\documentclass[particles,review,submit,moreauthors]{Definitions/mdpi} 
\usepackage{longtable}

\firstpage{1} 
\pubvolume{0}
\issuenum{0}
\articlenumber{0}
\copyrightyear{2026}
\externaleditor{} % More than 1 editor, please add `` and '' before the last editor name
\datereceived{ } 
\daterevised{ } % Comment out if no revised date
\dateaccepted{ } 
\datepublished{ } 
\usepackage{amsmath,amsfonts,amssymb,mathtools,cases,slashed,bm}
\usepackage{epsfig, graphicx}
\usepackage{color}
\usepackage[dvipsnames]{xcolor}
\usepackage{physics}
\usepackage{hhline}
\usepackage{mathrsfs}
\usepackage[normalem]{ulem}
\usepackage{multirow}
\usepackage{float}
\usepackage{appendix}
\usepackage[normalem]{ulem}

\usepackage{makecell}

\Title{Review 
 of Pion and Nucleon Gluon Parton Distribution Functions from Lattice QCD 
} 

\Author{{William Good} %MDPI: Please carefully check the accuracy of names and affiliations.%Author: Names are accurate
 $^{1,2,}${*} %MDPI: We added  marked the corresponding author according to those submitted online at susy.mdpi.com. Please confirm. %Author: we confirm this change
\orcidA{} and Huey-Wen Lin $^{1}${\orcidB{}} %MDPI: We changed \orcidA{} into \orcidB{}, please confirm. %Author: we confirm this change
 }

\address{%
$^{1}$ \quad Department of Physics and Astronomy, Michigan State University, East Lansing, MI 48824, USA; {hueywen@msu.edu} %MDPI: We added the email address here according to those submitted online at susy.mdpi.com. Please confirm. Same as below. %Author: this is the correct email address
\\
$^{2}$ \quad Department of Computational Mathematics,
  Science and Engineering, Michigan State University, \mbox{East Lansing, MI 48824, USA}}
\corres{Correspondence: {goodwil9@msu.edu}}
  
\abstract{
Gluons play an essential role in the strong interaction that binds quarks into hadrons, as described by quantum chromodynamics (QCD). 
Understanding how the fundamental dynamics of QCD give rise to hadron structure remains one of the outstanding challenges in nuclear and particle physics. 
Parton distribution functions (PDFs) provide a successful framework for describing high-energy scattering processes involving hadrons. 
However, our knowledge of gluon PDFs remains comparatively limited because of the experimental challenges associated with directly constraining the gluon structure of hadrons. 
In recent years, a major breakthrough has enabled the calculation of the Bjorken-$x$ dependence of PDFs directly from lattice QCD. 
Among the available approaches, the large-momentum effective theory (LaMET) and pseudo-PDF frameworks have emerged as the most widely used methods and have produced the majority of lattice determinations of parton distributions. 
These developments have sparked rapid progress in first-principles studies of the $x$-dependent structure of hadrons, including gluon PDFs of the nucleon and pseudoscalar mesons. 
In this review, we summarize recent advances in lattice QCD calculations of gluon PDFs {using the LaMET and pseudo-PDF methods}, discuss the current status and impacts of these developments, and highlight the remaining challenges and future opportunities.
}

\begin{document}
\section{Introduction}

%The gluon structure of hadrons
Quantum chromodynamics (QCD) describes the strong interaction as a non-Abelian
gauge theory of quarks and gluons, and~it is the gluons, the~self-interacting
carriers of the color force, that make the theory both rich and difficult. 
Although gluons are absent from the naive valence picture of hadrons, they dominate much of their internal structure: gluons carry roughly half of the momentum of a fast-moving nucleon, they proliferate at small momentum fraction $x$ through gluon radiation and recombination, and~their dynamics are responsible for confinement and for the bulk of the visible mass in the universe~\cite{Accardi:2012qut,Gross:2022hyw}. 
A quantitative, $x$-resolved knowledge of the gluon content of hadrons is therefore central to two of the most fundamental open questions in hadronic physics: the decomposition of the nucleon mass and spin in terms of the matrix elements of the QCD energy--momentum tensor~\cite{Ji:1994av,Jaffe:1989jz}. 
Because the gluon field strength does not couple directly to the electroweak probes used in deep-inelastic scattering (DIS), the~gluon distributions are among the least well-determined properties of hadrons, and~improving them for the nucleon and even more so for the pion has become a defining goal of the next generation of experimental facilities and theoretical tools~\cite{Accardi:2012qut,AbdulKhalek:2021gbh,Burkert:2022hjz,Proceedings:2026xrb}.

%Unpolarized gluon PDFs of the nucleon and the pion
At leading twist, the~unpolarized gluon parton distribution function (PDF) $g(x,Q^2)$ encodes the probability of finding a gluon carrying a longitudinal momentum fraction $x$ of the parent hadron. 
In the nucleon, the~gluon PDF is constrained only indirectly through the $Q^2$ evolution (scaling violations) of the quark structure functions and through gluon-initiated processes such as jet, heavy-quark, and~prompt-photon production, so it remains poorly known at both large $x$, where it falls steeply, and~small $x$, where it rises rapidly and may ultimately saturate~\cite{Hou:2019efy,Ball:2017nwa,AbdulKhalek:2021gbh,Achenbach:2023pba,Amoroso:2022eow,Ethier:2020way}. 
The situation is far more severe for the pion. 
As the pseudo-Goldstone boson of dynamical chiral symmetry breaking, the~pion is a keystone of emergent hadronic mass, yet its gluon content is almost unmapped: 
there are no free-pion targets, and~the available information stems from a handful of pion-induced Drell--Yan measurements and from leading-neutron electroproduction via the Sullivan process~\cite{Achenbach:2023pba,Arrington:2021biu,Aguilar:2019teb}. 
The planned Electron-Ion Collider (EIC), together with COMPASS++/AMBER at CERN and a proposed electron-ion collider in China, is projected to reduce the pion sea-quark and gluon PDF uncertainties by factors of $\sim$5--$10$ over most of the $x$ range and settle how the gluon content of the pion compares with that of the nucleon~\cite{Proceedings:2026xrb,Abir:2023fpo,Burkert:2022hjz,AbdulKhalek:2022hcn,AbdulKhalek:2021gbh,Accardi:2012qut}.

%Gluon helicity distribution and the proton spin}
When the nucleon is longitudinally polarized, the~gluon helicity distribution $\Delta g(x,Q^2)$ measures the net gluon polarization along the nucleon spin.
Determining this is a central objective in hadron structure physics because it quantifies the contribution of gluon spin to the overall spin of the proton. 
Although decades of polarized deep-inelastic scattering and polarized proton--proton collision experiments have established that quark spins account for only a fraction of the proton spin, the~gluon helicity distribution remains incompletely determined, particularly in the large-$x$ region, where experimental constraints are scarce. 
Similar to the broader challenge of unraveling the nucleon's internal structure in terms of its quark and gluon constituents, precise measurements over an extended kinematic range are essential for developing a quantitative understanding of QCD and the origin of nucleon spin.  
The central difficulty is kinematic: existing data reach only $x\gtrsim0.01$--$0.05$, so that the integral $\Delta G$ carries large uncertainties from the unmeasured small-$x$ region, leaving the resolution of the spin decomposition open~\cite{Borsa:2020lsz,AbdulKhalek:2021gbh}.
The EIC, as~the first polarized lepton--hadron collider, is designed to extend the coverage of $\Delta g(x)$ down to $x\sim10^{-4}$ and thereby to pin down $\Delta G$ and, by~inference, the~orbital angular momentum  contribution~\cite{Aschenauer:2015ata,AbdulKhalek:2021gbh}.
In particular, the~planned LHCspin experiment will uniquely probe the gluon helicity PDF in the large-$x$ region, where existing constraints are weakest, through spin-dependent measurements of heavy-flavor and quarkonium production in polarized fixed-target collisions, providing information that is complementary to both RHIC and the future EIC~\cite{LHCspin:2025lvj}.

%Lattice QCD and the $x$-dependence of gluon PDFs
Because PDFs are defined through light-cone correlation functions, they are not directly accessible in the Euclidean formulation of lattice QCD.
For decades, lattice calculations were therefore limited to the lowest few Mellin moments of PDFs before power-divergent operator mixing and the rapidly deteriorating signal-to-noise ratio became prohibitive~\mbox{\cite{Lin:2017snn,Ji:2020ect,Constantinou:2020hdm,Lin:2023kxn,Gross:2022hyw,Lin:2025hka}}. 
This longstanding limitation was overcome with the realization that matrix elements of highly boosted hadrons connected by spatially separated (nonlocal) operators can be factorized and matched onto the corresponding light-cone distributions. 
In this review, we focus on two complementary approaches that have emerged from this framework: the large-momentum effective theory (LaMET), or~``quasi-PDF'' approach~\cite{Ji:2013dva,Ji:2014gla,Ji:2020ect}, and~the ``pseudo-PDF'' method~\cite{Radyushkin:2017cyf}. These methods provide direct access to the full Bjorken-$x$ dependence of parton distributions from first principles.
Early applications at the physical pion mass successfully determined the isovector unpolarized, helicity, and~transversity quark PDFs, with~independent lattice formulations and methodologies yielding encouraging agreement with global fits~\cite{Alexandrou:2018pbm,Joo:2020spy,Lin:2018qky}. 
More recently, these techniques have been extended to the gluon sector. Exploratory calculations of gluon PDFs using the pseudo-PDF approach have been carried out for both the nucleon~\cite{Fan:2020cpa,HadStruc:2021wmh,Delmar:2023agv,Khan:2022vot} and pseudoscalar mesons~\cite{Fan:2021bcr,Salas-Chavira:2021wui,NieMiera:2025inn}, followed by physical-point and continuum-extrapolated studies~\cite{Fan:2022kcb,Fan:2022qve,Good:2023ecp}. 
On the LaMET side, the~first determinations of the nucleon gluon PDFs were only recently explored~\cite{Good:2024iur} and subsequently achieved~\cite{Good:2025daz}. Furthermore, calculations using multiple lattice spacings and ensembles, together with the self-renormalization framework in the hybrid scheme~\cite{Ji:2020brr,LatticePartonLPC:2021gpi}, have recently been reported by the MSULat and LPC collaborations~\cite{NieMiera:2025mwj,NieMiera:2025vcx,ChenChen:2025amm}. 
In parallel, promising studies have demonstrated that incorporating LaMET and pseudo-PDF lattice inputs into global QCD analyses can significantly improve the determination of gluon PDFs by combining first-principles calculations with experimental cross-section data~\cite{Good:2023ecp,Ablat:2025xzm,Karpie:2023nyg}.
These rapid developments have transformed lattice QCD from a tool capable of computing only a few PDF moments into one that can directly determine the $x$ dependence of gluon distributions, establishing lattice calculations as an increasingly important component of modern global PDF analyses and motivating the comprehensive review presented in this~article.

The remainder of this review is organized as follows. 
In Section~\ref{sec:Methodology}, we introduce the theoretical foundations of the two leading approaches for computing $x$-dependent parton distributions from lattice QCD, namely, the LaMET and pseudo-PDF formalisms. 
We discuss how each approach reconstructs the light-cone PDFs from Euclidean lattice observables, highlighting their conceptual differences, underlying connections, and~complementary features. 
In Section~\ref{sec:Results}, we summarize representative lattice QCD calculations of gluon PDFs obtained using these methods, followed by a discussion of their impact on global PDF analyses in Section~\ref{sec:Impacts}.
We then review ongoing developments and future directions in lattice QCD calculations of gluon PDFs, including applications of machine learning and theoretical advances aimed at substantially improving the signal-to-noise ratio of matrix elements at large Wilson-line separations. 
Finally, we conclude in Section~\ref{sec:conclusion} with an outlook on future opportunities and challenges in the~field.

%%%%%%%%%%%%%%%%%%%%%%%%%%%%%%%%%%%%%%%%%%%%%%%%%%%%%%%%%%%%%%%%%%
\section{Methodology}\label{sec:Methodology}
%%%%%%%%%%%%%%%%%%%%%%%%%%%%%%%%%%%%%%%%%%%%%%%%%%%%%%%%%%%%%%%%%%
%*Light cone gluon PDF definitions for unpolarized and helicity
\subsection{Theoretical~Framework}\label{sec:theory}
The unpolarized gluon PDF $g(x)$ and polarized gluon PDF $\Delta g(x)$ for a hadron $H$ are defined by the Fourier transform of light-front correlators~\cite{Collins:1989gx,Manohar:1990kr, Manohar:1990jx},
\begin{equation}\label{eq:LCPDF}
g(x,\mu^{2}) = \frac{1}{ 2 x P^+} \int_{-\infty}^\infty \frac{\mathrm{d} \xi^-}{2\pi} \, e^{-i x P^+ \xi^-} \langle H(P) | F_a^{+\mu}(\xi^-) 
 \mathcal{W}_{ab}(\xi^-, 0) \, F_{b,\mu}^+(0) | H(P) \rangle,
\end{equation}
{and} %MDPI: Format for same variables should be consistent through whole paper (italic/bold/superscript/uppercase, etc.), please check all and modify if not consistent. Same as all equations. %Author: We beleive that the format is consistent after one change of \mu -> \mu^2 in the equations directly above and below (this is not highlighted)
\begin{equation}\label{eq:LCPPDF}
\Delta g(x,\mu^{2}) = \frac{1}{ 2 x P^+} \int_{-\infty}^\infty \frac{\mathrm{d} \xi^-}{2\pi} \, e^{-i x P^+ \xi^-} \langle H(P,S) | F_a^{+\mu}(\xi^-) 
 \mathcal{W}_{ab}(\xi^-, 0) \, \widetilde{F}_{b,\mu}^+(0) | H(P,S) \rangle,
\end{equation}
where $|H(P)\rangle$ denotes an unpolarized hadron with momentum $P$, and~$|H(P,S)\rangle$ is a polarized hadron with polarization $S$. 
$F^{\mu\nu}_a = \partial^\mu A_a^\nu - \partial^\nu A_a^\mu - g_s f_{abc}A_b^\mu A_c^\nu$ is the gluon field strength tensor, with~$A^\mu_a$ being the gluon field, and~$\widetilde{F}_{a}^{\mu \nu} = \frac{1}{2}\epsilon_{\mu \nu \alpha \beta}F^{\alpha \beta}_a$ is the dual field strength tensor.
Light-front coordinates are given by $\xi^\pm=(\xi^t\pm \xi^z)/\sqrt{2}$.
Here, $x$ is the longitudinal momentum fraction carried by the gluons. 
We consider the PDFs to be renormalized in the $\overline{\rm MS}$ scheme, with~a scale $\mu$. 
Finally, gauge invariance is preserved by the insertion of a Wilson line,
\begin{align}
\mathcal{W}_{ab}(\xi^-, 0)= \mathcal{P}\,{\rm exp}\big[ -i g \int_0^{\xi^-} \mathrm{d} \eta^- \, A_c^+(\eta^-) t^c \big]_{ab},
\end{align}
where the indices $\{a,b\}$ are the color components in the adjoint representation of SU(3) and $t^c$ are the generators of the gauge group.
$\mathcal{P}$ denotes path~ordering.

The obvious problem with these definitions of PDFs for lattice QCD comes from the light-cone separation $\xi^-$. 
This issue can be addressed by recognizing that Euclidean correlators approach their light-cone counterparts in the short-distance or large-momentum limits.
These two limits motivate the pseudo-PDF~\cite{Radyushkin:2017cyf} and LaMET~\cite{Ji:2013dva,Ji:2014gla,Ji:2020ect} methods, respectively.
The short-distance and large-momentum limits do not commute with taking the ultraviolet (UV) cutoff of the theory to infinity, so both methods involve some perturbative matching between the Euclidean objects and the light-cone PDFs.
In particular, for~the unpolarized gluon PDF, there are several different multiplicatively renormalizable gluonic correlators which go to the same light-cone limit~\cite{Zhang:2018diq,Balitsky:2019krf}; however, all recent studies have focused on the operator
\begin{equation}\label{eq:up_gluon_op}
    O_g^\text{unpol.}(z) = F^{ti}(z)\mathcal{W}(z,0)F^{t}_{~i}(0) - F^{ij}(z)\mathcal{W}(z,0)F_{ij}(0),
\end{equation}
as it has a better signal-to-noise ratio and seems to converge to the light-cone PDF more quickly compared to others~\cite{Good:2024iur}.
Here, the~indices $i$ and $j$ sum over only the transverse ($x,y$) directions, and~$z$ is a spatial separation in the $z$ direction.
The preferred helicity gluon operator is structured similarly~\cite{Balitsky:2021cwr}:
\begin{equation}\label{eq:hel_gluon_op}
    O_g^\text{hel.}(z) = F^{ti}(z)\mathcal{W}(z,0)\widetilde{F}^{t}_{~i}(0) - F^{ij}(z)\mathcal{W}(z,0)\widetilde{F}_{ij}(0).
\end{equation}
{This} %MDPI: Please confirm if keep the no indentation format for paragraphs below the equations. please check all and confirm, same below. %Author: We confirm keeping no indentation
 operator, however, has some undesired but removable contamination away from the light-cone limit, which will be discussed more later in this section.
Both the pseudo-PDF and LaMET methods start with lattice matrix elements of these operators, sandwiched between hadron states, with~momentum $P_z$ in the $z$ direction:
\begin{equation}\label{eq:ME_def}
    h^B(z,P_z) = \langle H(P_z,S) | O_g(z)|H(P_z,S)\rangle,
\end{equation}
where $O_g(z)$ denotes either the unpolarized or helicity operator.
For the unpolarized matrix elements, we take a spin average, and~for the helicity, spin is taken in the $z$ direction.
There are several different ways to extract these matrix elements from lattice data, as~will be discussed in the next subsection.
The bare matrix elements are multiplicatively renormalizable as~\cite{Zhang:2018diq}
\begin{equation}
    h^B(z,P_z) = Z_L(1/a) e^{-|z|\delta m}h^R(z,P_z),
\end{equation}
where $Z_L(1/a)$ renormalizes the logarithmic UV divergences associated with the finite lattice spacing $a$, and~the exponential term renormalizes the linear divergence associated with the Wilson-line self-energy, $\delta m$.

Focusing first on the unpolarized gluon PDF in the pseudo-PDF method, the~bare matrix elements are typically renormalized via the double ratio scheme to construct the reduced pseudo-Ioffe-time distribution (RpITD):
\begin{equation}\label{eq:RpITD}
    \mathcal{M}(\nu=zP_z,z^2) = \frac{h^B(0,0)}{{h^B(0,P_z)}}\frac{h^B(z,P_z)}{h^B(z,0)},
\end{equation}
where division by $h^B(z,0)$ removes the renormalization factor, and~the $h^B(0,0)/h^B(z,0)$ term removes the kinematic factors.
Here, $\nu = zP_z$ is the dimensionless and Lorentz-invariant Ioffe time.
The RpITD can be matched perturbatively to the {$\overline{\text{MS}}$ scheme} gluon and quark singlet Ioffe-time distributions (ITDs) via
\vspace{-8pt}
\begin{adjustwidth}{-\extralength}{0cm}
\centering %% If there is a figure in wide page, please release command \centering, for Table, ``\textwidth" should be ``\fulllength"
\begin{align}\label{eq:pPDF-matching}
    \mathcal{M}(\nu, z^2) =& \int_0^1 du \frac{\mathcal{I}_g(u\nu, \mu^2)}{\mathcal{I}_g(0, \mu^2)}B_{gg}(u\nu,z^2\mu^2)~ + \nonumber \\
    &  \int_0^1 du \left[\frac{\mathcal{I}_\Sigma(u\nu, \mu^2)}{\mathcal{I}_g(0, \mu^2)}-\frac{{\mathcal{I}_\Sigma(0, \mu^2)}\mathcal{I}_g(\nu, \mu^2)}{\mathcal{I}_g(0, \mu^2)^2}\right]B_{gq}(u\nu,z^2\mu^2) + \mathcal{O}(z^2M_H^2, z^2\Lambda_\text{QCD}^2),
\end{align}
\end{adjustwidth}
where $B_{gg}$ and $B_{gq}$ are the {perturbative next-to-leading order (NLO)} gluon--gluon and gluon--quark matching kernels described in Refs.~\cite{Balitsky:2019krf, Balitsky:2021qsr}{. 
We note} %MDPI: Note is not allowed for this journal, so we changed it into content in maintext. Please confirm. Same as notes below. %Author: We changed the parentheticals to normal sentences, and in some cases added a few extra words to maintain flow
 {that some terms were neglected in Ref.}~\cite{Balitsky:2019krf}{ and were corrected in the more complete calculation provided in Ref.}~\cite{Balitsky:2021qsr}.
Here, $\mu^2$ is the $\overline{\text{MS}}$ scheme renormalization scale, and~the ITDs are defined by the Fourier transform of the gluon and quark singlet PDFs
\begin{equation}
    \mathcal{I}_g(\nu, \mu^2) = \frac{1}{2}\int_{-1}^1dx~e^{ix\nu} xg(x,\mu^2)
\end{equation}
and
\begin{equation}
    \mathcal{I}_\Sigma(\nu, \mu^2) = \frac{1}{2}\int_{-1}^1dx~e^{ix\nu} x\Sigma(x,\mu^2).
\end{equation}
It should be noted that the ITDs at zero are equal to the PDF moments: $\mathcal{I}_i(0,\mu^2) = \langle x_i\rangle (\mu^2)$.
{Mixing with the quark arises from operator mixing in the $\overline{\text{MS}}$ scheme. 
The complicated non-perturbative structure in front of $B_{gq}$ is due to the normalization $\mathcal{M}(0,z^2) = 1$ in the ratio scheme. 
Both $B_{gg}(0,\mu^2) = B_{gq}(0,\mu^2) = 1$, which makes it clear that the perturbative matching relation is consistent with this normalization.
In many cases mixing is ignored, as~its effect has been found to be negligible compared to current statistical uncertainties in pseudo-PDF studies~\cite{Fan:2021bcr, Fan:2022kcb,Delmar:2023agv}.}
This matching relation has higher-twist effects that come in at $\mathcal{O}(z^2M_H^2, z^2\Lambda_\text{QCD}^2)$, which means that only short-distance data should be used in the matching expression to obtain the PDF.
{It is worth mentioning that normalization of the RpITD means that only $xg(x)/\langle x_g\rangle$ can be extracted from the methodology as laid out.
The momentum fractions can be extracted from independent studies of local operators on the same lattice ensembles.
The extracted gluon momentum fractions still exhibit substantial variation across studies, and~the uncertainties are much larger than those of phenomenological determinations, which are constrained by the sum rule $\sum_i \langle x_i \rangle = 1$.}

{To obtain a light-cone PDF from lattice data via the pseudo-PDF method, one takes the lattice matrix elements described by Equation~(\ref{eq:ME_def}) at discrete values of the lattice momentum $P_z\in[0,P_z^\text{max}]$ at $z=0$ and one or more discrete values of $z>0$.
The RpITD, $\mathcal{M}(\nu=zP_z,z^2)$, is constructed via the definition in Equation~(\ref{eq:RpITD}) at discrete values of $\nu$ up to some $\nu^\text{max}$ for each $z$.}
In some studies~\cite{HadStruc:2021wmh, Delmar:2023agv}, Equation~(\ref{eq:pPDF-matching}) is inverted to obtain the ITD directly { at the same discrete values of $\nu$, which also requires some interpolation of the RpITD between values of $\nu$.
Because data for the ITD exist only up to some value $\nu^\text{max}$, there is an inverse problem in completing the inverse Fourier transform of the gluon ITD to obtain $xg(x,\mu^2)$.
The common solution for gluon PDFs is to constrain parameters for a simple phenomenological model of the PDF, such as $x^\alpha (1-x)^\beta$, using a $\chi^2$ fit between the ITD produced from the Fourier transform of the model PDF and the lattice ITD.
For quark PDFs obtained via the pseudo-PDF method, more sophisticated methods, including Bayesian and machine learning extractions, have been developed and applied~\cite{Chowdhury:2024ymm,Dutrieux:2024rem,Dutrieux:2025jed}.
On the other hand,} if we consider the cosine transform of the $B_{gg}$ and $B_{gq}$ kernels, we can construct matching relations directly between the RpITD and PDFs, as~described in Ref.~\cite{Balitsky:2019krf}, {sidestepping the need for interpolation.}
A similar fit can be performed by comparing the RpITDs produced by the model with the lattice RpITDs, as~is performed by the MSULat group~\mbox{\cite{Fan:2020cpa, Fan:2021bcr, Fan:2022kcb, Salas-Chavira:2021wui, NieMiera:2025inn}}.
{Bayesian and machine learning methodologies have not been widely explored for this direct-matching relation, as~the direct matching is more computationally expensive than evaluating the Fourier transforms.
The one- and two-step matching methods are mathematically equivalent, but~with real lattice data the~interpolation step may introduce some bias.
The tradeoff between these two methods could be further explored.}
{It is important to appreciate that the pseudo-PDF method does not intrinsically impose an unreliable range of $x$; rather, the~reliability and proper estimation of the reliability depend on the method used to solve the inverse problem.
Recent studies using Bayesian methods and synthetic data out to values of $\nu^\text{max}=4$ suggest} that the factorization provides constraints throughout the entire $x\gtrsim 0.2$ region of the PDF~\cite{Medrano:2025cmg}.

The process for the helicity gluon PDF is similar overall but includes a few caveats.
The RpITD is written
\begin{equation}\label{eq:hel_RpITD}
    \widetilde{\mathcal{M}}(\nu,z^2) = \frac{M_H^2}{P_zP_t}\frac{\widetilde{h}^B(z,P_z)}{h^B(z,0)Z_L(z/a)},
\end{equation}
where $M_H$ is the mass of the hadron in question, $P_t$ is the temporal momentum, and~$Z_L(z/a)$ is an extra factor introduced to cancel UV logarithms; it is described in more detail in Ref.~\cite{Balitsky:2021cwr}.
This RpITD contains a known $\mathcal{O}(M_H^2/P_z^2)$ contamination term, which can be seen by comparing Lorentz decompositions of the Euclidean and light-cone matrix elements~\cite{Balitsky:2021cwr}. 
Several methods have been proposed and applied to subtract this contamination~\cite{HadStruc:2022yaw,Khan:2022vot,Chowdhury:2024ymm}, allowing the use of the cleaner, low-momentum data.
From the contamination-subtracted RpITD, the~rest of the analysis is analogous to that described for the unpolarized PDF, where the kernels for matching described in Equation~(\ref{eq:pPDF-matching}) are replaced by the expressions provided in Ref.~\cite{Balitsky:2021cwr}; however, current extractions are still too noisy to obtain PDFs directly from the lattice~RpITD. 

The process in LaMET is somewhat different.
At large separations, double-ratio renormalization can introduce non-perturbative effects~\cite{Ji:2020brr}. 
This is not an issue in the pseudo-PDF method because only short-distance matrix elements should be used; however, long-distance data are required in the LaMET method, so a more sophisticated renormalization scheme is necessary.
In particular, hybrid-ratio renormalization is typically used in LaMET calculations of the gluon PDF; this is defined as follows:
\begin{equation} \label{eq:renormalized-quasi-LF}
    h^{R}(z, P_{z})
    =
    N\frac{h^{B}(z, P_{z}, 1/a)}{h^{B}(z, P_{z}=0, 1/a)} \, \theta(z_{s} - |z|) 
    +  T_{s} \, \frac{h^{B}(z, P_{z}, 1/a)}{Z_{R}(z, 1/a)} \, \theta(|z| - z_{s}),
\end{equation}
Here, $z_s$ is a cutoff length chosen below approximately $0.3-0.4$ fm, beyond~which perturbation theory becomes unreliable.
The factors $N$ and $T_s$ are for normalizing, which we usually choose to fix the matrix elements to $1$ at $z=0$ and to ensure continuity at $z=z_s$, respectively. 
At short distances, ratio renormalization is used, as~can be seen by the division by the zero-momentum matrix elements. 
When only one lattice spacing is available, the~$Z_{R}(z,1/a)$ factor is taken just to be $e^{{-}(\delta m + m_0)|z|}$, where $m_0$ is the renormalon ambiguity.
This arises in the perturbative calculation for the zero-momentum matrix elements, called the Wilson coefficients, which are defined by
\begin{align}\label{eq:wil-coef}
C_0(z,\mu) = 1 + \frac{\alpha_s C_A}{2\pi} \left( \frac{5}{6} \ln \left( \frac{z^2 \mu^2 e^{2\gamma_E}}{4} \right) + \frac{3}{2} \right).
\end{align}
With one lattice spacing, only the sum $\delta m + m_0$ can be extracted.
The sum, $\delta m + m_0$, is then calculated by fitting the natural log of the ratio between the zero-momentum lattice matrix elements and the Wilson coefficient
\begin{equation}\label{eq:dm_fit}
(\delta m + m_0) z  + I_0 \approx \ln\left[ \frac{{C_0}(z,\mu)}{h^{\mathrm{B}}(z,0)} \right],
\end{equation}
where $I_0$ denotes a $z$-independent constant. 
In more recent studies with multiple lattice spacings, the~self-renormalization scheme~\cite{LatticePartonLPC:2021gpi} has been used, which involves the extraction of $\delta m$ and $m_0$ independently, along with logarithmic terms, lattice spacing dependence, and~residual non-perturbative effects from the zero-momentum matrix elements on different ensembles.
More details on the exact process are given in recent gluon LaMET papers~\cite{NieMiera:2025mwj,ChenChen:2025amm, NieMiera:2025vcx}.

The Fourier transform of the renormalized matrix elements defines what is called the quasi-PDF:
\begin{equation}
    x\widetilde{g}(x,P_z) = \frac{P_z}{(P_t)^2}\int \frac{dz}{2\pi}e^{izxP_z}h^R(z,P_z).
\end{equation}
In practice, the~lattice data at large separation distances are too noisy to perform a direct Fourier transform.
However, it is expected that the matrix elements decay exponentially at long distance, so in modern calculations, physically inspired fit forms~\cite{Ji:2020brr,Gao:2021dbh,Ji:2026vir} are used to extrapolate the matrix elements out to infinity.
Theoretical bounds on the systematic uncertainty associated with these extrapolations have been derived as well~\cite{Gao:2021dbh,Ji:2026vir}.
The quasi-PDF can be matched to the light-cone PDF via a double Fourier transform of \mbox{Equation~(\ref{eq:pPDF-matching})} plus a correction term to match between the hybrid-ratio or self-renormalization schemes and the $\overline{\text{MS}}$ renormalization scheme.
This is typically written as
\vspace{-10pt}
\begin{adjustwidth}{-\extralength}{0cm}
\centering %% If there is a figure in wide page, please release command \centering, for Table, ``\textwidth" should be ``\fulllength"
\begin{align}\label{eq:mtcheq}
 x\tilde{g}\left(x, P_z\right)= & \int_{-1}^1 \frac{d y}{|y|} C_{gg}\left(\frac{x}{y}, \frac{\mu}{y P_z}\right)  y g(y, \mu^2) + C_{gq}\left(\frac{x}{y}, \frac{\mu}{y P_z}\right)\Sigma(y,\mu^2) + \mathcal{O}\left(\frac{\Lambda_\text{QCD}^2}{x^2(1-x)^2P_z^2}\right).
\end{align}
\end{adjustwidth}
The exact kernel for $C_{gg}$ is written in Refs.~\cite{Good:2025daz}, while the mixing terms have been ignored so far.
{It has not been explored whether this term is negligible in the LaMET case given current uncertainties.
Similar to the pseudo-PDF methodology, the~normalization of the matrix elements also constrains this method to only being able to calculate $xg(x)/\langle x_g \rangle$.}
The LaMET method has the advantage that when the long-distance behavior of the matrix elements can be reliably extrapolated, it provides a fully $x$-dependent PDF.
However, the~expansion breaks down at large- and small-$x$ {when $\frac{\Lambda_\text{QCD}}{x(1-x)P_z}>1$}.
Because of the poor signal-to-noise ratio at large separations and the additional contamination term, the~helicity gluon PDF has not yet been extracted via the LaMET method, but~the main methodological flow will be analogous when these two issues are~resolved.

\subsection{{Extraction of Lattice Matrix~Elements}}
In lattice QCD, the~extraction of the nonlocal hadronic matrix elements defined by Equation~(\ref{eq:ME_def}) typically comes from the analysis of two-point and three-point  correlators.
The two-point correlators are defined as
\begin{equation} \label{eq:2pt-correlator}
    C^{\text{2pt}}_H (P_z; t_\text{sep}) = \int d^3 y \; e^{-i y P_z} \langle 0 | \chi_H (\vec{y}, t_\text{sep})  \chi^\dagger_H (\vec{0},0) | 0 \rangle,
\end{equation}
where $t_\text{sep}$ is the Euclidean separation time, $P_z$ is the hadron boost momentum in the z-direction, and~$\chi^\dagger(y)$  is the interpolating operator for the given hadron.
These interpolators create and annihilate a tower of states with the same quantum numbers as the hadron of interest.
The textbook choices for the nucleon and pion interpolators are $\chi^\dagger_N = \epsilon_{abc}(d_a^T C \gamma_5 u_b)\mathcal{P}_+ u_c$ and  $\chi^\dagger_\pi=\overline{u}\gamma_t\gamma_5 d$.
Here, $C$ denotes the charge-conjugation matrix and $\mathcal{P}_+$ projects onto positive-parity states
New interpolators with better signal-to-noise ratios have been proposed recently in Refs.~\cite{Zhang:2025hyo,Reitinger:2026hta}, but~not yet widely applied to the gluon PDF calculations.
Another technique commonly used to improve the signal at large momentum is Gaussian momentum smearing~\cite{Bali:2016lva}.
The three-point correlators are defined by the insertion of the nonlocal gluon operator:
\begin{equation} \label{eq:3pt-correlator}
    C^{\text{3pt}}_H(P_z,  z; t_\text{sep}, t) = 
    \int d^3 \; y e^{-iyP_z} \langle 0 | \chi_H (\vec{y}, t_\text{sep}) O_g(z,t) \chi_H^\dagger(\vec{0}, 0) | 0 \rangle,
\end{equation}
where $t_\text{sep}$ is the separation time, and~$t$ is the insertion time of the operator.
The gluon operator only couples to the quarks via disconnected diagrams, meaning that the three-point correlators %EE: Please check meaning retained
can be constructed explicitly by measuring many two-point correlators and combining them with the vacuum measurements of $O_g(z,t)$:
\begin{equation}
    \langle C^\text{3pt}_H(P_z,  z; t_\text{sep}, t)\rangle = \langle C^{\text{2pt}}_H (P_z; t_\text{sep}) O_g(z,t) \rangle - \langle C^{\text{2pt}}_H (P_z; t_\text{sep})\rangle \langle O_g(z,t) \rangle.
\end{equation}
This is in contrast to connected quark diagrams, where the quark operators have to be constructed with explicit contractions with the propagators between the source and sink.
Disconnected contributions have substantially poorer signal-to-noise ratios than connected contributions, as~they are more sensitive to UV~fluctuations.

The two-point correlator has an eigenvalue expansion as follows:
\begin{equation} \label{eq:2pt-correlator-exp}
    C^{\text{2pt}}_H (P_z; t_\text{sep}) = \sum_{n=0}^\infty |A_n|^2 e^{-E_nt_\text{sep}},
\end{equation}
where $A_n$ represents overlap factors between the $n$th state of the hadron and the interpolators and $E_n$ is the energy of the $n$th state.
The three-point correlator has a similar expansion:
\begin{equation} \label{eq:3pt-correlator-exp}
    C^{\text{3pt}}_H (P_z, z; t_\text{sep}, t) = \sum_{n,m=0}^\infty A_n A_m^* \langle H^{(m)}(P_z)|O_g(z)| H^{(n)}(P_z) \rangle e^{-E_m(t_\text{sep}-t)}e^{-E_nt},
\end{equation}
where $|H^{(n)}(P_z) \rangle$ is the $n$th excited state of the hadron.
At large values of $t_\text{sep}$, assuming that excited-state contamination is small, Equations~(\ref{eq:2pt-correlator-exp}) and~(\ref{eq:3pt-correlator-exp}) can be truncated to only include the first one or two states.
The simplest methodology to extract matrix elements is by fitting a plateau of the ratio between the three-point and two-point correlators at large $t_\text{sep}$, as~\begin{equation} \label{eq:2pt-3pt-correlator-ratio}
    R(P_z,z;t_\text{sep}, t) = \frac{C^{\text{3pt}} (P_z, z; t_\text{sep}, t)}{C^{\text{2pt}} (P_z; t_\text{sep})} = h(z,P_z) + \mathcal{O}(e^{-(E_1-E_0) t_\text{sep}}).
\end{equation}
In calculations of the gluon, excited-state contamination causes this ratio to converge slowly, and~in certain pathological cases this can give incorrect results at finite values of $t_\text{sep}$~\cite{Hackett:2024xnx}.
An obvious method to handle this is to truncate the spectral expression in Equations~(\ref{eq:2pt-correlator-exp}) and~(\ref{eq:3pt-correlator-exp}) to $n,m = 2$ or $3$ and fit the resulting models.
$A_{n,m}$, $E_{n,m}$, and~$\langle H^{(m)}(P_z)|O_g(z)| H^{(n)}(P_z) \rangle$ are taken as parameters, fitted by minimizing the $\chi^2$ between lattice correlator data and the models.
This can be performed one at a time, starting with the two-point fit, using the overlaps and energies as priors or fixed values for the three-point fit, or~one can simultaneously fit the two-point and three-point data. %EE: Please check meaning retained
Another method involves constructing summed operators or simply summing the ratios in Equation~(\ref{eq:2pt-3pt-correlator-ratio}) over the insertion time $t$.
The slope of this sum will approach the desired matrix element at large $t_\text{sep}$~\cite{Maiani:1987by,Dong:1997xr,Bulava:2011yz}.
Finally, the extraction of matrix elements can be recast as a question of extracting eigenvalues and eigenvectors. 
The most well-known treatment is the aptly named generalized eigenvalue problem (GEVP)~\cite{Edwards:2011jj}.
This method can also be combined with the summed operators method described before~\cite{Bulava:2011yz, Blossier:2009kd}.
A more recent proposal is the use of the Lanczos algorithm~\cite{Lanczos1950,Wagman:2024rid,Hackett:2024xnx,Hackett:2024nbe}.
Reference~\cite{Abbott:2025yhm} explores the relation of Lanczos and GEVP with many other proposed eigenvalue methods for spectroscopy and matrix-element extraction and serves as a good source to explore a variety of these~methods.

\begin{figure}[H]
    
    \includegraphics[width=0.425\textwidth]{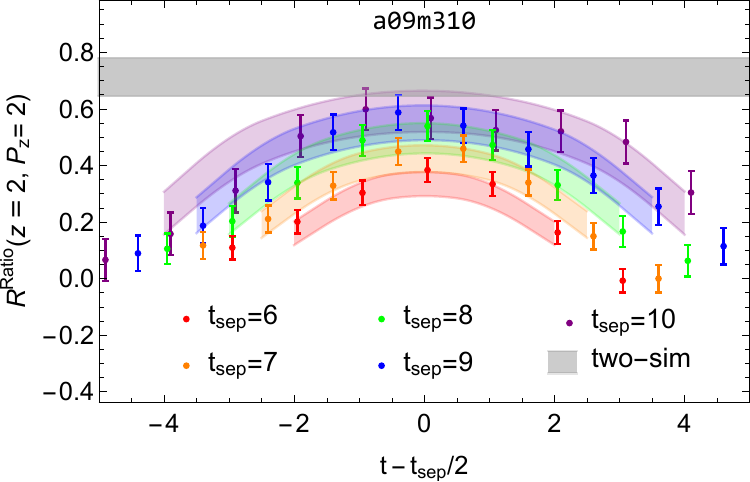}
    \includegraphics[width=0.25\textwidth]{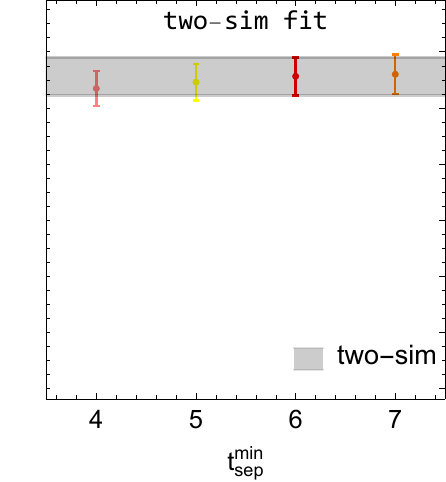}
    \includegraphics[width=0.25\textwidth]{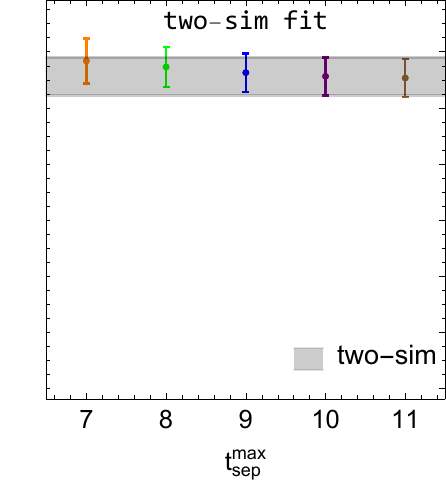}
    \caption{{Example} %MDPI: Please note that changes to the position/size of figures or tables may occur during the production stage.
    %MDPI: Please confirm whether an explanation of the colors on the right images needs to be added to the figure caption.%Author: The colors are just for visual distinction of the different points, which are fully described by their values on the x-axis. We do not think this warrants explicit explanation
 of stability and fit-quality plots for two-state matrix-element fits from Ref.~\cite{Fan:2022kcb}.
    The gray band shows the extracted value for the ground-state matrix element.
    The left plot shows the ratio from Equation~(\ref{eq:2pt-3pt-correlator-ratio}) as a function of $t-t_\text{sep}/2$, together with the reconstruction obtained from the fit to the data with $t_\text{sep}/a \in[6,10]$.
    The middle and right plots show extracted matrix elements when varying the ranges: $t_\text{sep}/a \in[t_\text{sep}^\text{min},10]$ and $[6,t_\text{sep}^\text{max}]$, respectively.
    $z$ and $P_z$ in the label are in integer lattice units.}
    \label{fig:ratio_plot}
\end{figure}

Three-point correlators of gluonic operators suffer from significant excited-state contamination, so two- or multi-state fits and carefully handled eigenvalue methods should be regarded as more reliable than plateau fits.
For example, we show a figure visualizing the two-state-fit quality from an MSULat study of the unpolarized nucleon gluon PDF~\cite{Fan:2022kcb} in \mbox{Figure~\ref{fig:ratio_plot}}.
On the leftmost panel, the~ratio defined in Equation~(\ref{eq:2pt-3pt-correlator-ratio}) is plotted for both the lattice data and the bands reconstructed from the two-state fit of the lattice data in $t_\text{sep}/a \in [6,10]$ to the models defined by truncating \mbox{Equations~(\ref{eq:2pt-correlator-exp}) and (\ref{eq:3pt-correlator-exp})} to $n,m = 2$.
The gray band represents the extracted value of $h(z,P_z) = \langle N^{(0)}(P_z)|O_g(z)| N^{(0)}(P_z) \rangle$.
As one can see, the~ratio converges slowly to the extracted matrix elements, so a plateau fit to the ratio at any single value $t_\text{sep}$ would underestimate the matrix element.
In the center and right plots, the~fits are recomputed with data in varying ranges: $t_\text{sep}/a \in[t_\text{sep}^\text{min},10]$ and $[6,t_\text{sep}^\text{max}]$, respectively.
This demonstrates control of excited-state contamination, as~there is convergence as $t_\text{sep}^\text{min}$ and $t_\text{sep}^\text{max}$ increase.
That is, neglecting the contributions from higher-order states with $n,m \geq 3$ at finite $t_\text{sep}$ does not affect the extraction of the matrix~elements.

\unskip

\subsection{{Discussion of Sources of~Uncertainty}}
Uncertainty in LaMET and pseudo-PDF calculations comes from a variety of sources, which can be separated into statistical and systematic uncertainties.
Statistical uncertainty arises from the fact that lattice calculations involve Monte Carlo sampling of observables on top of vacuum gauge configurations.
This is worse in the gluon case due to the fact that gluonic contributions come only from noisy disconnected diagrams, as discussed in the previous section.
Systematic uncertainties can be subdivided into multiple categories, including lattice uncertainties, matching (perturbative and higher-twist) uncertainties, and~epistemic uncertainties associated with the various choices made in the analysis process.
The independent variables associated with the lattice ensembles and kinematics of the initial matrix elements interact in ways that affect both statistical precision and systematic uncertainties. 
It is important not to conflate how various independent variables and analysis choices increase and decrease different statistical and systematic uncertainties depending on the~methodology.

Lattice artifacts arise inherently from the discretization of QCD in a finite box of volume $V = L^3\times T$ with lattice spacing $a$.
Lattice-spacing dependence is often taken to be $\mathcal{O}(a^n)$ with $n=1$ or $2${. This choice of $n$ comes from the fact that most modern lattice actions are ``$\mathcal{O}(a)$-improved'', meaning that the discretized action differs from the continuum starting at $\mathcal{O}(a^2)$; however,
this only limits measured quantities to a maximum expected accuracy of $\mathcal{O}(a^2)$. Discretized operators and discrete operations (e.g.,~Fourier transforms) on those quantities may be accurate only up to $\mathcal{O}(a)$}.
The coefficient will depend on the choice of lattice action.
Finite-volume effects are often ignored in gluon PDF calculations as chiral effective theory demonstrates that they are exponentially suppressed by a factor of $\exp[-M_\pi L]$~\cite{Colangelo:2010ba}.
Most modern lattice calculations have $M_\pi L \gtrsim 3$ and often $M_\pi L \gtrsim 4$, so uncertainties here are at the percent level.
Although not strictly required, lattice calculations are often performed at non-physical quark masses, which reduces statistical noise and speeds up the computation at the cost of some potential bias in the physics. 
Extrapolations to remove the quark mass dependence are usually parameterized by $(M_\pi^\text{lat.} - M_\pi^\text{phys.})^n$, with~$n=1$ or $2$~\cite{Fan:2022kcb, ChenChen:2025amm}.
Additionally, the~effects associated with setting the number of flavors of sea quarks to at least $N_f = 2+1$ should be small in the gluon compared to other sources of uncertainty.
One more important lattice-related systematic is gauge link smearing, utilized heavily in gluon calculations to smooth out UV fluctuations, reducing statistical noise.
Too much smearing can bias the results.
Different studies use different smearing methods, including hypercubic (HYP)~\cite{Hasenfratz:2001hp}, stout~\cite{Morningstar:2003gk}, and~Wilson flow \cite{Luscher:2010iy}{. Wilson flow is also called gradient flow when used as a renormalization scheme}~\cite{Brambilla:2023vwm}{. We will use the former throughout this review}.
HYP and stout smearing are implemented as a finite number of discrete steps, while Wilson flow is implemented as a continuous ``flow-time'' variable $\tau$, which has units of length squared.
There have not been any detailed comparisons between the different smearing methods and different amounts of smearing.
The kinematic variables of hadron momentum, $P_z$, and~operator separation distance, $z$, both represented in Equation~(\ref{eq:ME_def}), give rise to unique systematics.
Both kinematic variables interact with lattice-spacing effects, as~discretization effects are expected to grow at large momentum and shrink at longer distances.
These effects can be parameterized by $aP_z$ and $a/z$, respectively~\cite{Karpie:2021pap, JeffersonLabAngularMomentumJAM:2022aix, ChenChen:2025amm}.
Most of the traditional lattice artifacts are reasonably understood. 
Ref.~\cite{FlavourLatticeAveragingGroup:2019iem} gives many details about these artifacts and lattice actions, for~example.
Smearing effects and the interplay of discretization effects with kinematic variables still need to be understood~better.

Of course, the~kinematic variables are well known to contribute to the higher-twist effects in the LaMET and pseudo-PDF methods, as~mentioned before with the $\mathcal{O}(\frac{\Lambda^2_\text{QCD}}{x^2(1-x)^2P_z^2})$ and $\mathcal{O}(z^2\Lambda^2_\text{QCD}, z^2M_H^2)$ errors in the matching.
The treatment of higher-twist effects as a systematic uncertainty depends on the factorization framework used to relate Euclidean observables to light-cone PDFs.
That is, the~effects are introduced because the matching relations in Equations~(\ref{eq:pPDF-matching}) and~(\ref{eq:mtcheq}) are only accurate up to leading twist.
Some quark and phenomenological studies have attempted to remove higher-twist effects in the pseudo-PDF method by parameterization~\cite{Karpie:2021pap, JeffersonLabAngularMomentumJAM:2022aix, ChenChen:2025amm, Good:2025nny}, but~this remains to be explored in lattice-only gluon studies.
On the other hand, state-of-the-art LaMET studies are limited to boost momenta of 3 GeV or less. 
Conservatively, results are considered reliable within $x\in[0.2,0.8]$ or smaller via the criteria $\frac{\Lambda_\text{QCD}}{x(1-x)P_z}>1$.
These estimates are subjective, as~$\Lambda_\text{QCD}$ is a parameter typically chosen to be in the range of $300-500$ MeV.
Much more work could be performed here to better quantify the reliable region with considerations of higher-twist effects.
Truncation of the perturbative matching calculation to finite order introduces an additional systematic uncertainty.
Gluon LaMET and pseudo-PDF matching has only reached NLO accuracy~\cite{Wang:2017qyg, Wang:2017eel, Wang:2019tgg,Balitsky:2019krf, Balitsky:2021qsr, Balitsky:2021cwr, Yao:2022vtp, Ma:2022gty, Ji:2025cbb, Monahan:2026skv}; this could be improved by completing higher loop calculations.
Another direction for improvement in perturbative matching is to handle renormalization group resummation~\cite{Su:2022fiu}, leading renormalon resummation~\cite{Zhang:2023bxs}, and~threshold resummation~\cite{Ji:2023pba,Liu:2023onm,Ji:2024hit}, as has been performed in the quark case.
These improvements together act to reduce uncertainty due to scale dependence in the perturbative matching for LaMET.
Similar improvements have not been discussed widely in the pseudo-PDF case. 
Matching systematics, as~a whole, are controllable and are incrementally improved as more studies reach higher momentum and smaller lattice spacings; however, they could be better quantified in both LaMET and pseudo-PDF~studies.

Finally, epistemic uncertainties arise due to a variety of methodological choices in the analysis pipeline, which makes these uncertainties difficult to quantify in the final results.
The sources of epistemic uncertainties discussed most often include the method used to solve the inverse problem in the pseudo-PDF case and the extrapolation models in the LaMET case.
Both uncertainties arise because inverse Fourier transforms (or inverse matching) must be computed from a limited set of data.
As discussed in Section~\ref{sec:theory}, gluon pseudo-PDF studies have been limited to quite simple models and,~as such, more sophisticated methods need to be explored~\cite{Chowdhury:2024ymm,Dutrieux:2024rem,Dutrieux:2025jed}.
Obtaining larger-momentum data also reduces this uncertainty, as~the large-$\nu$ constraints make the inverse problem less ill-posed.
In the LaMET case, the~recent asymptotic forms should only introduce uncertainty at the level of ``just a few percent''~\cite{Ji:2026vir}, but~large-$z$ data is quite noisy, and~these new extrapolation forms have not been widely explored yet.
The choice to ignore quark mixing, which is made in many of the gluon PDF studies, also brings uncertainty.
Mixing comes in only at NLO and is smaller than the purely gluonic NLO matching contribution because of the color factor $C_F$ versus $C_A$, so it is generally small compared to other sources of statistical and systematic error.
Future precision studies must address this, though.
Another potential source of epistemic uncertainty comes from matrix-element extraction.
As discussed in the previous section, extraction of the matrix elements from correlators is non-trivial and, if~treated improperly, can give rise to excited-state contamination in the results.
However, carefully checked two-state fits and eigenvalue methods~\cite{Edwards:2011jj,Bulava:2011yz, Blossier:2009kd,Lanczos1950,Wagman:2024rid,Hackett:2024xnx,Hackett:2024nbe,Abbott:2025yhm} are regarded as quite reliable. 
Aside from these widely discussed sources of epistemic uncertainty, the different renormalization methods covered in Section~\ref{sec:theory} have not been compared directly, and~the two-step versus one-step matching methods in the pseudo-PDF method may cause differences in the results.
The one-step method may be preferable because it avoids interpolation-related bias.
Finally, the~choice of how the $\mathcal{O}(M^2_H/P_z^2)$ contamination is handled in the helicity gluon PDF could be a source of epistemic uncertainty and is a major hurdle that has not been fully tackled.
Overall, although~epistemic uncertainties are difficult to quantify, the~two most concerning sources in the two methods have been tackled in the quark case and, as~such, should be tacked in the gluon case in upcoming studies as large-$z$ and large-$P_z$ precision increases and helps to rule out poor model choices.
Other sources are regarded as small or under~control.

To summarize, statistical and systematic uncertainties need to be carefully separated.
The independent variables associated with the lattice setup and kinematics give rise to uncertainties in multiple categories.
It is perhaps most informative to give a concrete example.
The lattice momentum, $P_z$, is associated with poorer statistical precision and $\mathcal{O}(aP_z)$ ``momentum discretization'' effects. 
However, $P_z$ suppresses higher-twist effects in LaMET, increasing the reliable range of LaMET PDFs. 
Pseudo-PDF factorization is reliable at all values of $P_z$, but~having more precise large-$P_z$ improves the precision of inverse-problem solutions~\cite{Dutrieux:2025jed}. 
Finally, large-$P_z$ lattice correlator data often have more excited-state contributions, which can contaminate the extracted matrix elements if not handled carefully.
This exemplifies the challenges faced in attempting to simultaneously improve the statistical and systematic precision in both methodologies.
Despite these challenges, considerable effort has been devoted to understanding and reducing statistical and systematic uncertainties in lattice determinations of gluon PDFs from the lattice, as~will be discussed in the next~section.

\section{Gluon PDF~Results}\label{sec:Results}
In this section, we will cover the history of unpolarized and polarized gluon PDFs extracted from the Euclidean correlators defined in Equations~(\ref{eq:up_gluon_op})--(\ref{eq:ME_def}).
The main challenge in extracting gluon PDFs is the poor signal-to-noise ratio in the disconnected diagrams needed to extract the ground-state hadronic matrix elements of the operators.
Considerable effort has been devoted to obtaining the first gluon-PDF results and subsequently reducing their systematic uncertainties.
Table~\ref{tab:lattice_gluon_pdf} summarizes all the full calculations of unpolarized gluon PDFs for the pion and proton, along with the first published work toward obtaining a helicity gluon PDF in the proton.
These gluon PDF studies use a variety of different lattice spacings, pion masses, fermion actions, and~number of flavors, which give rise to many different competing statistical and systematic effects, which need to be considered when comparing the results of these studies.
Many of the studies use relatively coarse lattice spacings compared to state-of-the-art quark calculations~\cite{Gao:2021dbh,Gao:2023ktu, Gao:2026hix, Grebe:2026qmt}, and~the pion masses are nearly all heavier than physical.
Only a handful of studies~\cite{Fan:2022kcb, NieMiera:2025mwj, ChenChen:2025amm} have attempted continuum extrapolations, and~only two studies have extrapolated to the physical pion mass~\cite{Fan:2022kcb, ChenChen:2025amm}.
All studies use a significant amount of gauge link smearing compared to quark studies, which commonly only have one step of HYP.
Refs.~\cite{HadStruc:2021wmh, HadStruc:2022yaw} attempt to remove the smearing effects with a zero-flow-time extrapolation, and~in Ref.~\cite{NieMiera:2025vcx} it is suggested that self-renormalization seems to remove the effects of Wilson flow comfortably within one standard deviation when either $\tau$ or $\tau/a^2$ is taken to be a constant. 
In the next subsection, we will cover the unpolarized gluon PDF studies in more~detail.

\begin{table}[H]
%\centering
\caption{{Summary} %MDPI: We added inside horizontal lines for the table, please confirm. %Author: We're okay with this change 
 of lattice calculations of gluon PDFs.
The first column gives the reference number for each study. The~PDF column describes which PDF (unpolarized or polarized) and which hadron is being studied. The~lattice-details column provides four pieces of information for each study, describing the fermion action, number of flavors $N_f$, approximate lattice spacing(s) $a$, and~approximate pion mass(es) $M_\pi$. The~method column states whether the pseudo-PDF or LaMET method was used and which renormalization scheme is used. The~final column describes the type of gauge link smearing used and the number of steps or flow times.}
\label{tab:lattice_gluon_pdf}

\begin{adjustwidth}{-\extralength}{0cm}
%\centering %% If there is a figure in wide page, please release command \centering, for Table, ``\textwidth" should be ``\fulllength"

\begin{tabularx}{\fulllength}{L C L C C}
\toprule
\textbf{Reference}
& \textbf{PDF}
& \makecell[c]{\textbf{Lattice Details}}
& \textbf{Method}
& %Gauge 
   \textbf{Link~Smearing} \\
\midrule

MSULat'20~\cite{Fan:2020cpa}
&
$g^p(x)/\langle x^p\rangle$
&
%\makecell[l]{
%\multicolumn{1}{l}{
Clover-to-HISQ  \linebreak
$N_f=2+1+1$\linebreak
$a\approx0.12~\mathrm{fm}$\linebreak
$M_\pi\approx310\text{--}690~\mathrm{MeV}$
%}
&
{Pseudo-PDF\linebreak Double ratio}
&
{HYP\linebreak 5 steps}
\\
\midrule

MSULat'21~\cite{Fan:2021bcr}
&
$g^\pi(x)/\langle x^\pi\rangle$
&
%\makecell[l]{
Clover-to-HISQ\linebreak 
$N_f=2+1+1$\linebreak 
$a\approx0.12\text{--}0.15~\mathrm{fm}$\linebreak 
$M_\pi\approx220\text{--}690~\mathrm{MeV}$
%}
&
{Pseudo-PDF\linebreak Double ratio}
&
{HYP\linebreak 5 steps}
\\
\midrule

HadStruc'21~\cite{HadStruc:2021wmh}
&
$g^p(x)\langle x^p\rangle$
&
%\makecell[l]{
Wilson-clover\linebreak 
$N_f=2+1$\linebreak 
$a\approx0.09~\mathrm{fm}$\linebreak 
$M_\pi\approx360~\mathrm{MeV}$
%}
&
{Pseudo-PDF\linebreak Double ratio}
&
{Wilson flow\linebreak $\tau/a^2=1\text{--}3.8$}
\\
\midrule

MSULat'22~\cite{Fan:2022kcb}
&
$g^p(x)$
&
%\makecell[l]{
Clover-to-HISQ\linebreak 
$N_f=2+1+1$\linebreak 
$a\approx0.09\text{--}0.15~\mathrm{fm}$\linebreak 
$M_\pi\approx220\text{--}690~\mathrm{MeV}$
%}
&
{Pseudo-PDF\linebreak Double ratio}
&
{HYP\linebreak 5 steps}
\\
\midrule

HadStruc'22~\cite{HadStruc:2022yaw}
&
%$\Delta g^p(x)$
$\langle x^p\rangle  \widetilde{\mathcal{M}}(\nu,z^2)$
&
%\makecell[l]{
Wilson-clover\linebreak 
$N_f=2+1$\linebreak 
$a\approx0.09~\mathrm{fm}$\linebreak 
$M_\pi\approx360~\mathrm{MeV}$
%}
&
{Pseudo-PDF\linebreak Modified single ratio}
&
{Wilson flow\linebreak  $\tau/a^2=1\text{--}3.8$}\\

\bottomrule

\end{tabularx}

\end{adjustwidth}
\end{table}

\begin{table}[H]\ContinuedFloat
%\centering
\caption{\textit{Cont.}}
\label{tab:lattice_gluon_pdf}

\begin{adjustwidth}{-\extralength}{0cm}
%\centering %% If there is a figure in wide page, please release command \centering, for Table, ``\textwidth" should be ``\fulllength"

\begin{tabularx}{\fulllength}{L C L C C}
\toprule
\textbf{Reference}
& \textbf{PDF}
& \makecell[c]{\textbf{Lattice Details}}
& \textbf{Method}
& %Gauge 
   \textbf{Link~Smearing} \\
\midrule

%MSULat'20~\cite{Fan:2020cpa}
%&
%$g^p(x)/\langle x^p\rangle$
%&
%%\makecell[l]{
%%\multicolumn{1}{l}{
%Clover-to-HISQ  \linebreak
%$N_f=2+1+1$\linebreak
%$a\approx0.12~\mathrm{fm}$\linebreak
%$M_\pi\approx310\text{--}690~\mathrm{MeV}$
%%}
%&
%{pseudo-PDF\linebreak Double ratio}
%&
%{HYP\linebreak 5 steps}
%\\
%\midrule
%
%MSULat'21~\cite{Fan:2021bcr}
%&
%$g^\pi(x)/\langle x^\pi\rangle$
%&
%%\makecell[l]{
%Clover-to-HISQ\linebreak 
%$N_f=2+1+1$\linebreak 
%$a\approx0.12\text{--}0.15~\mathrm{fm}$\linebreak 
%$M_\pi\approx220\text{--}690~\mathrm{MeV}$
%%}
%&
%{pseudo-PDF\linebreak Double ratio}
%&
%{HYP\linebreak 5 steps}
%\\
%\midrule
%
%HadStruc'21~\cite{HadStruc:2021wmh}
%&
%$g^p(x)\langle x^p\rangle$
%&
%%\makecell[l]{
%Wilson-clover\linebreak 
%$N_f=2+1$\linebreak 
%$a\approx0.09~\mathrm{fm}$\linebreak 
%$M_\pi\approx360~\mathrm{MeV}$
%%}
%&
%{pseudo-PDF\linebreak Double ratio}
%&
%{Wilson flow\linebreak $\tau/a^2=1\text{--}3.8$}
%\\
%\midrule
%
%MSULat'22~\cite{Fan:2022kcb}
%&
%$g^p(x)$
%&
%%\makecell[l]{
%Clover-to-HISQ\linebreak 
%$N_f=2+1+1$\linebreak 
%$a\approx0.09\text{--}0.15~\mathrm{fm}$\linebreak 
%$M_\pi\approx220\text{--}690~\mathrm{MeV}$
%%}
%&
%{pseudo-PDF\linebreak Double ratio}
%&
%{HYP\linebreak 5 steps}
%\\
%\midrule
%
%HadStruc'22~\cite{HadStruc:2022yaw}
%&
%%$\Delta g^p(x)$
%$\langle x^p\rangle  \widetilde{\mathcal{M}}(\nu,z^2)$
%&
%%\makecell[l]{
%Wilson-clover\linebreak 
%$N_f=2+1$\linebreak 
%$a\approx0.09~\mathrm{fm}$\linebreak 
%$M_\pi\approx360~\mathrm{MeV}$
%%}
%&
%{pseudo-PDF\linebreak Modified single ratio}
%&
%{Wilson flow\linebreak  $\tau/a^2=1\text{--}3.8$}\\

%\midrule
%
ETMC'23~\cite{Delmar:2023agv}
&
$g^p(x)$
&
%\makecell[l]{
Twisted-mass\linebreak 
$N_f=2+1+1$\linebreak 
$a\approx0.09~\mathrm{fm}$\linebreak 
$M_\pi\approx260~\mathrm{MeV}$
%}
&
{Pseudo-PDF\linebreak Double ratio}
&
{Stout\linebreak 10--20 steps}
\\
\midrule

MSULat'24a~\cite{Good:2023ecp}
&
$g^\pi(x)$
&
%\makecell[l]{
Clover-to-HISQ\linebreak 
$N_f=2+1+1$\linebreak 
$a\approx0.12\text{--}0.15~\mathrm{fm}$\linebreak 
$M_\pi\approx220\text{--}690~\mathrm{MeV}$
%}
&
{Pseudo-PDF\linebreak Double ratio}
&
{HYP\linebreak 5 steps}
\\
\midrule

MSULat'24b~\cite{Good:2024iur}
&
$g^p(x)/\langle x^p\rangle$
&
%\makecell[l]{
Clover-to-HISQ\linebreak 
$N_f=2+1+1$\linebreak 
$a\approx0.12~\mathrm{fm}$\linebreak 
$M_\pi\approx310\text{--}690~\mathrm{MeV}$
%}
&
{LaMET\linebreak  “Ansatz”  hybrid \linebreak ratio}
&
{HYP and Wilson\linebreak 5 steps and $\tau/a^2=3$}
\\
\midrule

MSULat'25a~\cite{Good:2025daz}
&
$g^p(x)/\langle x^p\rangle$
&
%\makecell[l]{
Clover-to-HISQ\linebreak 
$N_f=2+1+1$\linebreak 
$a\approx0.12~\mathrm{fm}$\linebreak 
$M_\pi\approx310\text{--}690~\mathrm{MeV}$
%}
&
{LaMET\linebreak Hybrid-ratio}
&
{HYP and Wilson\linebreak 5 steps and $\tau/a^2=3$}
\\
\midrule

MSULat'25b~\cite{NieMiera:2025mwj}
&
$g^p(x)/\langle x^p\rangle$
&
%\makecell[l]{
Clover-to-HISQ\linebreak 
$N_f=2+1+1$\linebreak 
$a\approx0.09\text{--}0.15~\mathrm{fm}$\linebreak 
$M_\pi\approx310~\mathrm{MeV}$
%}
&
{LaMET\linebreak Hybrid-Self}
&
{Wilson flow\linebreak $\tau/a^2=3$}
\\
\midrule

LPC'25~\cite{ChenChen:2025amm}
&
$g^p(x)/\langle x^p\rangle$
&
%\makecell[l]{
Wilson-clover\linebreak 
$N_f=2+1$\linebreak 
$a\approx0.05\text{--}0.11~\mathrm{fm}$\linebreak 
$M_\pi\approx135\text{--}320~\mathrm{MeV}$
%}
&
{LaMET\linebreak Hybrid-Self}
&
{HYP\linebreak 10 steps}
\\
\midrule

MSULat'25c~\cite{NieMiera:2025vcx}
&
$g^p(x)/\langle x^p\rangle$
&
%\makecell[l]{
Clover-to-HISQ\linebreak 
$N_f=2+1+1$\linebreak 
$a\approx0.09\text{--}0.15~\mathrm{fm}$\linebreak 
$M_\pi\approx310\text{--}690~\mathrm{MeV}$
%}
&
{LaMET\linebreak Hybrid-Self}
&
{Wilson flow\linebreak $\tau/a^2=1\text{--}5$}
\\

\bottomrule

\end{tabularx}

\end{adjustwidth}
\end{table}

\subsection{Unpolarized}
\begin{figure}[H]
    
    \includegraphics[width=\linewidth]{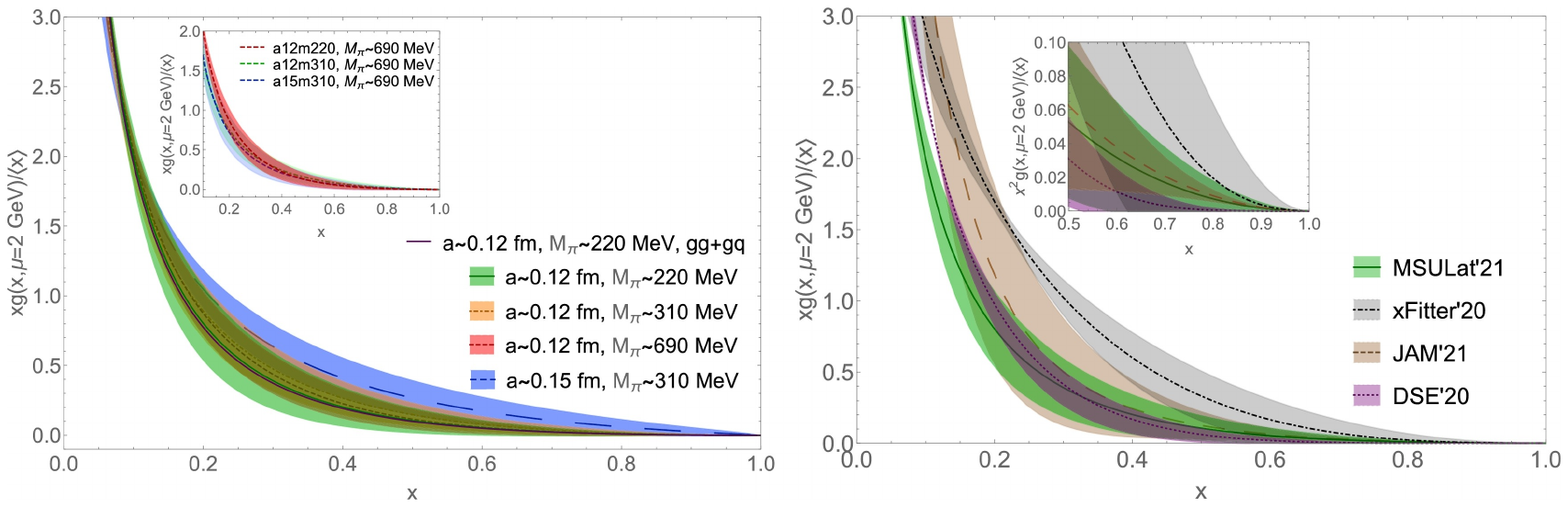}
    \caption{{\textbf{(Left)}} %MDPI: please confirm if it is the explanation for subfigure on left, if yes, please add the explanation for right subfigure. %Author: Clarified which part of the sentence was referring to the left vs right plots
    Comparison of the extracted pion gluon PDF across several lattice spacings and {\textbf{(right)}} of the $a\approx0.12$~fm, $M_\pi\approx220$~MeV PDF with phenomenological results from JAM~\cite{Barry:2018ort}, xFitter~\cite{Novikov:2020snp}, and~an extraction based on the Dyson–Schwinger equation~\cite{Cui:2020tdf}. Figure taken from MSULat'21~\cite{Fan:2021bcr}. 
    }
    \label{fig:pion-PDF}
\end{figure}

The first attempt to study nonlocal gluon operators with the goal of extracting a gluon PDF was Ref.~\cite{Fan:2018dxu}; however, this was extremely preliminary and demonstrated that the signal-to-noise ratio of the operator was very poor.
Additionally, they used a nonlocal generalization of the gluonic energy--momentum tensor, which turned out not to be multiplicatively renormalizable~\cite{Wang:2019tgg}.
Calculations using either LaMET or the pseudo-PDF method were not feasible until the authors of Ref.~\cite{Balitsky:2019krf} defined the new operator in Equation~(\ref{eq:up_gluon_op}), which has a much better signal-to-noise ratio~\cite{Good:2024iur} and fewer higher-twist effects than other previously proposed operators.
The first pseudo-PDF studies using this operator began to appear soon after for both the proton~\cite{Fan:2020cpa,HadStruc:2021wmh, Fan:2022kcb, Delmar:2023agv} and~the pion~\cite{Fan:2021bcr}.
Starting with the pion, the study in~\cite{Fan:2021bcr} was performed with the typical MSULat setup, shown in Table~\ref{tab:lattice_gluon_pdf}, focusing on the two larger lattice spacings $a\approx 0.12$ and $0.15$ fm.
The main results for this study are shown in Figure~\ref{fig:pion-PDF}.
The left plot shows that the results from different lattice spacings and pion masses are all statistically consistent, possibly indicating that the extracted PDF depends more strongly on lattice spacing than on pion mass. 
On the right, it can be seen that the lattice result is consistent with results from the Dyson--Schwinger equation (DSE)~\cite{Cui:2020tdf} and JAM~\cite{Barry:2018ort} at $x\gtrsim0.2$, while xFitter~\cite{Novikov:2020snp} is in some tension with the other results.
{Different phenomenological extractions likely differ because of the lack of gluon data in the pion, and~although early lattice results favor the JAM and DSE determinations, the~coarse lattice spacing and non-physical pion mass prevent strong conclusions from being made.
There have not been substantial updates on the gluon PDF in the pion since this early work, aside from the inclusion of lattice data into global fits, which will be discussed in Section~\ref{sec:Impacts}.
Additional studies at higher statistical and systematic precision are certainly needed in the pion case.}

\begin{figure}[H]
    
    \includegraphics[width=0.475\linewidth]{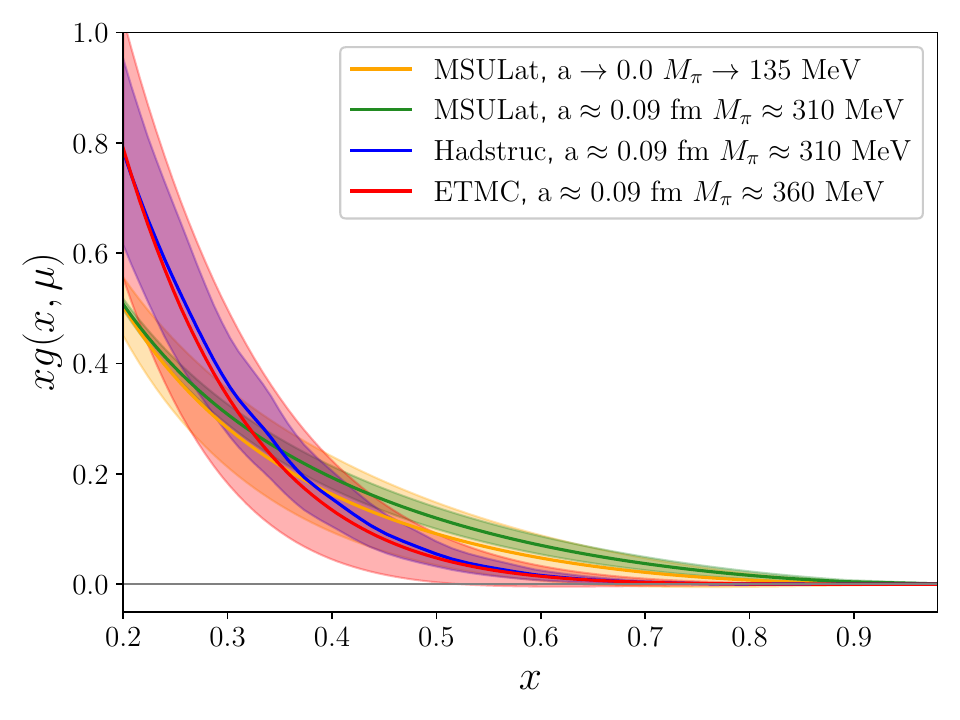}
    \includegraphics[width=0.475\linewidth]{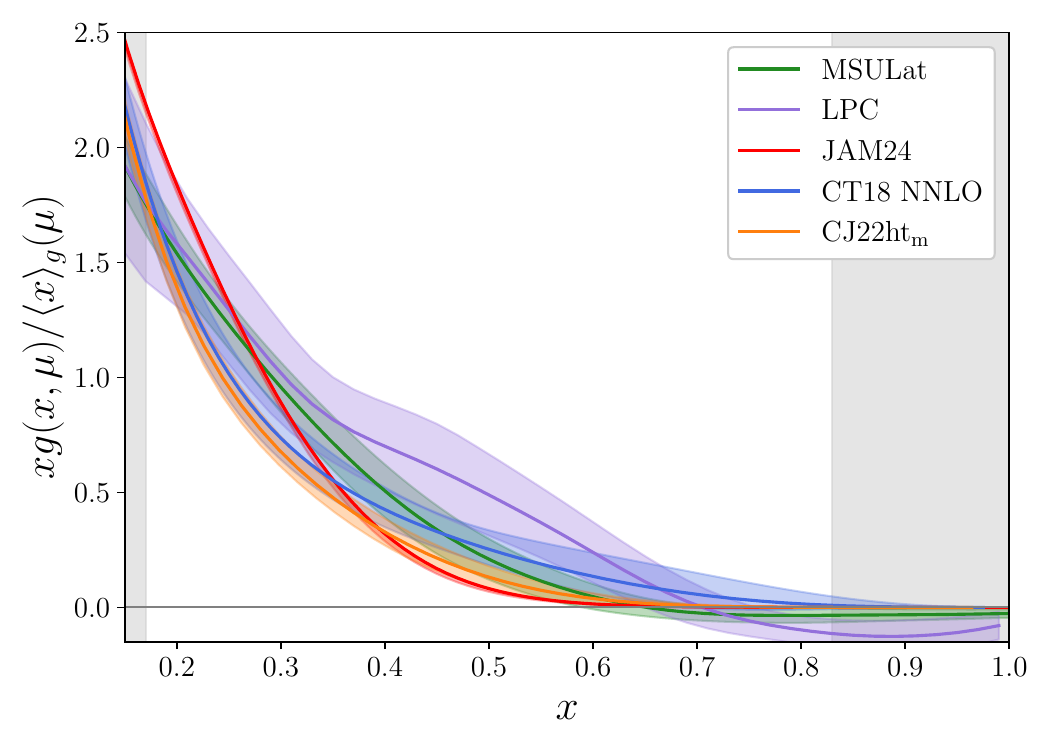}
    \caption{{(\textbf{Left})} %MDPI: Please confirm whether an explanation of the colored shaded area needs to be added to the figure caption.%Author: Added explanation below
 Comparison of the gluon PDFs obtained using the pseudo-PDF method from MSULat at the physical-continuum limit and at $a\approx 0.09$~fm~\cite{Fan:2022kcb}, HadStruc~\cite{HadStruc:2021wmh}, and~ETMC~\cite{Delmar:2023agv}.
({\textbf{Right})}~Comparison of the MSULat continuum-extrapolated gluon PDF from the LaMET method with the LPC continuum- and physical-limit-extrapolated PDF, along with phenomenological PDFs from CT18~\cite{Hou:2019efy}, JAM24~\cite{Anderson:2024evk}, and~CJ22~\cite{Cerutti:2025yji}.
{The shaded regions represent where the LaMET matching breaks down for the MSULat PDF.}}
    \label{fig:PDF-comp}
\end{figure}

Focusing on the three most recent nucleon studies, which correspond to the most up-to-date analyses from the HadStruc %EE: Please check meaning retained
\cite{HadStruc:2021wmh}, Extended Twisted Mass (ETMC)~\cite{Delmar:2023agv}, and~MSULat~\cite{Fan:2022kcb} collaborations, it is important to note that the three different studies each involve choices which affect the systematics of the final PDFs.
Each group uses different fermion actions, numbers of flavors, lattice spacings, and~pion masses, as described in the third column of Table~\ref{tab:lattice_gluon_pdf}.
{Each study uses a different matrix-element extraction method, with~ETMC opting for the older plateau fit methodology, MSULat using two-state fits, and~HadStruc %EE: Please check meaning retained
using the summed GEVP method.}
HadStruc %EE: Please check meaning retained
and ETMC match the RpITD to the ITD, then fit a parameterization of the PDF using the ITD data, while MSULat fits the PDF parameters directly from the matching between the PDF and RpITD. 
{It is unknown whether this distinction has a significant effect on the results.}
The fermion actions and number of flavors likely make a small impact on the results of gluonic quantities, but~the pion mass and lattice spacing could still affect the results. 
An important note is that MSULat uses three lattice spacings and three pion masses to complete a physical-continuum extrapolation.
Additionally, each group uses different gauge link smearing techniques, which are not directly comparable.
HadStruc %EE: Please check meaning retained
attempts to extrapolate to zero flow time, which should remove the effects.
The ETMC study is also the only study to use lattice data to handle the quark mixing terms in the matching, shown in Equation~(\ref{eq:pPDF-matching}), but~{{Figure 16} %MDPI: There's no Figure 16 in this paper, please check if the citation is correct. if it indicates figure 16 in Reference study?%Author: This is in reference to the figure in Ref. [33]. Changed "their study" to Ref. [33] to clarify this
 of {Ref.}~\cite{Delmar:2023agv} shows the effect of mixing to be much smaller than statistical uncertainties.}
With all of these systematics in mind, we compare the pseudo-PDF studies alongside several phenomenological fits on the left side of Figure~\ref{fig:PDF-comp}.
Here, the~MSULat physical continuum and $a\approx 0.09$ fm $M_\pi \approx 310$ MeV PDFs are both shown.
From this plot, it can be seen that the HadStruc %EE: Please check meaning retained
and ETMC gluon PDFs are highly consistent with each other, while the MSULat PDFs are slightly larger above $x\gtrsim 0.5$.
A possible reason behind this tension is that the HadStruc %EE: Please check meaning retained
and ETMC fits constrain $\alpha \geq 0$ in the $x^\alpha(1-x)^\beta$ parametrization, while the MSULat PDF does not apply such a constraint.
{The choice $\alpha>0$ forces integrability of $g(x)$, which is not a necessary physical constraint.
On the other hand, $\alpha > -1$ is a necessary constraint, which ensures the integrability of $xg(x)$, hence a finite $\langle x_g\rangle$.
The impact of this constraint has not been fully explored, as~these studies do not actually quote their extracted values of $\alpha$.
In any case, using more modern non-parametric approaches~\cite{Dutrieux:2024rem} could reduce dependence on restrictive parametrization assumptions, and~should be strongly considered for future pseudo-PDF extraction of the gluon PDF.
This is likely one of the dominant systematic uncertainties that need to be tackled. 
Future studies should focus as well on lattice systematics at higher statistical precision.
}

On the LaMET side, MSULat completed an initial study of ratio- and hybrid-ratio-renormalized matrix elements for LaMET from several different operators in Ref.~\cite{Good:2024iur}.
This study brought to the attention of the community that LaMET matching formulas needed to be derived for the operator in Equation~(\ref{eq:up_gluon_op}), as~the signal at large Wilson-line separations had become sufficiently precise as to obtain a quasi-PDF for this operator, while the other operators had much worse signal.
This quickly resulted in collaboration to derive the appropriate matching kernel, and~the first gluon PDF from LaMET was calculated in Ref.~\cite{Good:2025daz} using hybrid-ratio renormalization, at~a fixed lattice spacing of $a\approx 0.12$ fm at pion masses of $M_\pi \approx 310$ and $690$ MeV.
This study showed that pion mass dependence was small, while the smearing dependence was significant.
After this, MSULat~\cite{NieMiera:2025mwj}, and later the Lattice Parton Collaboration (LPC), each produced continuum-extrapolated gluon PDFs with self-renormalization~\cite{ChenChen:2025amm}.
LPC also added a chiral extrapolation to their study and an infinite-momentum extrapolation.
An important note of these studies is that MSULat used a fixed relative flow time of $\tau/a^2 = 3$ for the gauge link smearing, while LPC used 10 steps of HYP smearing on all ensembles.
We show the results of these two studies compared with phenomenological gluon PDFs on the right side of Figure~\ref{fig:PDF-comp}.
The gray bands show where LaMET is generally agreed to be unreliable for boost momentum, around $P_z\approx 2$ GeV.
The shapes of the LPC and MSULat PDFs are not in qualitative agreement, but~the large errors on the LPC PDF, which include careful estimations of systematic errors, leave the two PDFs within 1.5 standard deviations across the entire range.
The LPC PDF becomes highly negative, outside of $1\sigma$, more so than the MSULat PDF, which is a phenomenologically undesirable behavior{, as~this PDF should be interpretable as probability density.}
The PDFs from both studies generally are in $1\sigma$ agreement with the CT18 PDF in the intermediate-$x$ regions, and~both PDFs show some tension with the JAM24 and CJ22 analyses below around $x=0.4\text{--}0.5$.
Overall, it is difficult to draw definitive conclusions from the results here, as~there are still systematics that need to be controlled better.
{In particular, although~LPC has made great effort in reducing their lattice systematics with finer lattice spacings and pion masses closer to the physical point, some of these ensembles, although~better suited for controlling systematic effects, have not been measured with equally high statistical precision.
Additionally, their choice of single-state matrix-element fits with summed operators could be extended to two-state fits or eigenvalue methods.
Nonetheless, this study is a major step forward in controlling lattice systematics.}

\begin{figure}[H]
    
    \includegraphics[width=0.475\linewidth]{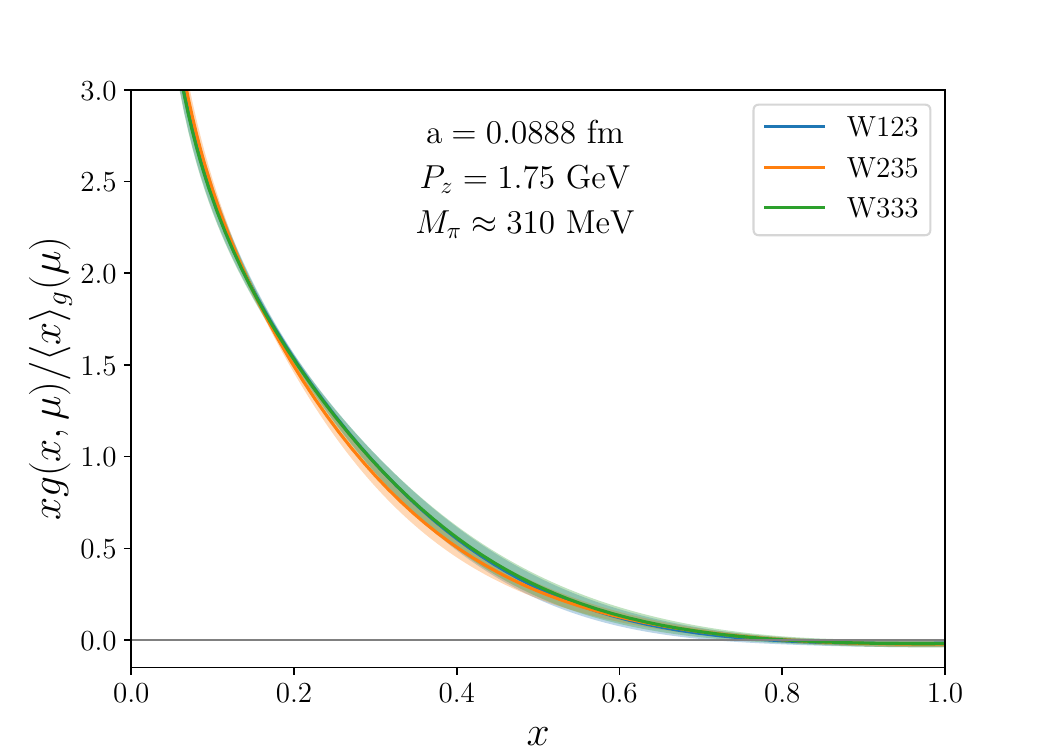}
    \includegraphics[width=0.45\linewidth]{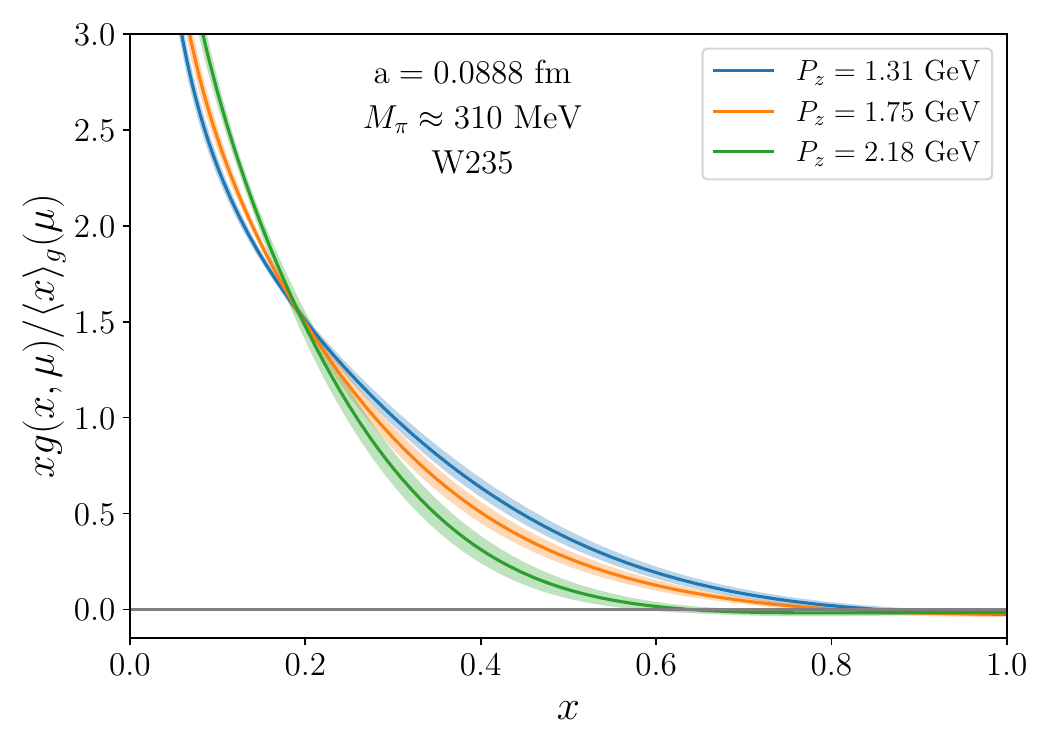}
    \caption{The dependence on Wilson flow time used in the self-renormalization study (\textbf{left}) and the dependence of the final PDFs on the boost momentum (\textbf{right}). 
    Figures taken from MSULat'25c~\cite{NieMiera:2025vcx}.}
    \label{fig:systematic_study}
\end{figure}

In that vein, MSULat paired their shorter continuum extrapolation study with a broader systematic study~\cite{NieMiera:2025vcx}.
Two important findings of this study are that the momentum dependence in the extracted PDFs is still quite large and,~unlike in the hybrid-ratio renormalization case, the~smearing dependence seems to be largely removed by self-renormalization, as~long as either a fixed relative $\tau/a^2$ or fixed physical $\tau$ flow time is used.
Two sample plots show these two claims in Figure~\ref{fig:systematic_study}.
On the left, W123, W235, and~W333 represent using $\tau/a^2 = \{1,2,3\},\{2,3,5\},\{3,3,3\}$ for the $a\approx\{0.15, 0.12, 0.09\}$ fm lattice spacings.
W123 and W333 both have $\tau/a^2 = 3$ for the $a\approx0.09$ fm ensemble, but~the agreement between these two plots demonstrates that differences in smearing choices for the other ensembles do not affect the ability of self-renormalization to properly renormalize the matrix elements, giving consistent results.
The W235 results demonstrate that the extracted PDF at a different fixed flow time is statistically consistent with those from the W123 analysis.
{In Ref.~\cite{NieMiera:2025vcx}, the~heuristic explanation behind why fixed $\tau/a^2$ versus fixed $\tau$ does not change the results is that Wilson flow changes the linear divergence in the bare matrix elements by an additive factor proportional to $1/\sqrt{8\tau}$.
Both methods contribute to modulations of the linear divergence that can be absorbed and renormalized by the self-renormalization scheme.
This was not an intentional design of self-renormalization, and~needs to be studied more deeply, as~it could be used to intentionally build handling of smearing into self-renormalization or other renormalization techniques.}
On the right, three different momenta are used in the analysis, and~there are very significant differences in the extracted PDFs, indicating that either perturbative matching has not sufficiently converged, momentum-discretization effects are large, or~higher-twist effects remain significant.
The origin of this tension should be explored further and resolved.
{MSULat still needs to explore smaller pion masses and push to lower lattice spacings, as~LPC has, but~this study identifies momentum convergence as one of the most pertinent issues to explore.}
\vspace{-6pt}

\unskip

\subsection{Helicity}
\begin{figure}[H]
    
    \includegraphics[width=0.45\linewidth]{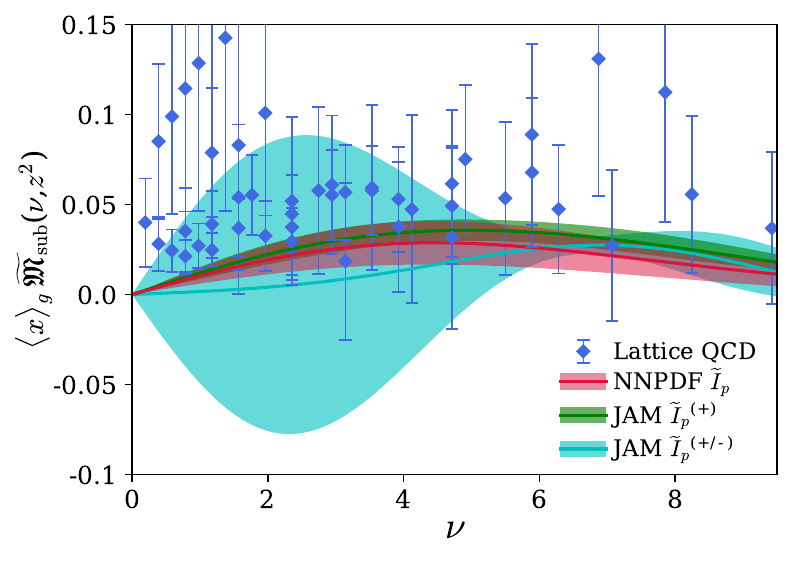}
    \includegraphics[width=0.45\linewidth]{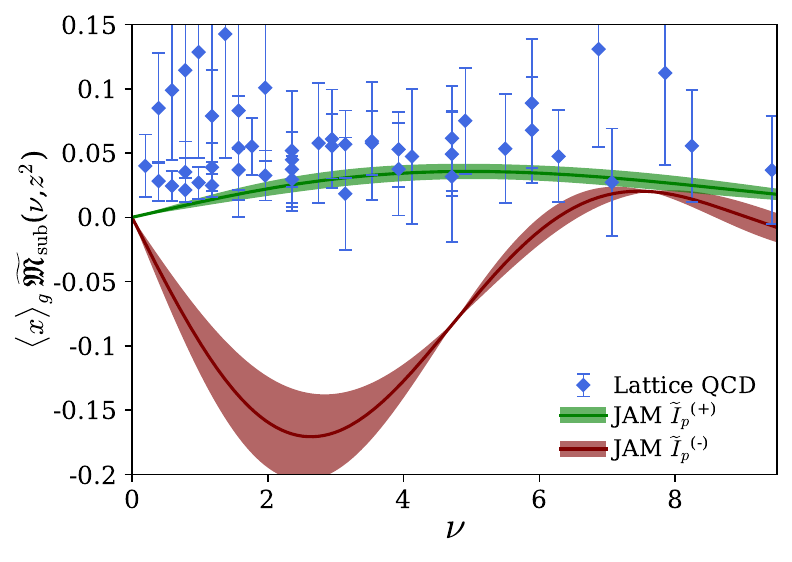}
    \caption{{Comparison} %MDPI: Please change the hyphen (-) into a minus sign (−, “U+2212”) in the figure, e.g., “-1” should be “−1”.%Author: This figure comes from Ref. [61], and we do not have the original data or code that generated these plots.
 of HadStruc’s contamination-subtracted RpITDs (blue points) with the ITDs reconstructed from the NNPDF~\cite{Nocera:2014gqa} and JAM~\cite{Zhou:2022wzm} global fits. In~the (\textbf{left}) plot, the~cyan band corresponds to the JAM fit allowing $\Delta G$ to be either positive or negative, while the green band includes only the JAM solutions with positive $\Delta G$. In~the (\textbf{right}) plot, the~green band is the same as that in the left plot, while the maroon band corresponds to the solutions with negative $\Delta G$. Plots taken from HadStruc'22~\cite{HadStruc:2022yaw}.}
    \label{fig:hadstruc_pol}
\end{figure}

To date, there has been only one published lattice extraction of the gluon helicity RpITD~\cite{HadStruc:2022yaw}, coming from the HadStruc collaboration, where the results were not used to fit a final gluon PDF.
This study used the same ensemble as their unpolarized gluon PDF in Ref.~\cite{HadStruc:2021wmh}, which was described in the previous sections.
They again extracted matrix elements with the summed GEVP method at several values of Wilson flow time and extrapolated to zero flow time.
In this study, particular care was taken to extract the $\mathcal{O}(M^2_H/P_z^2)$ contamination term mentioned in Section~\ref{sec:Methodology}.
They explored a fit to the Mellin moments of the contaminating distribution, along with constructing a subtraction scheme using the zero-momentum matrix elements, as~is described further in Ref.~\cite{HadStruc:2022yaw}.
They find consistency between the two extraction methods.
One of HadStruc's %EE: Please check meaning retained
goals was to help resolve the sign of the gluon helicity contribution:
\begin{equation}
\Delta G(\mu^2) = \int_0^1dx~\Delta g(x,\mu^2).    
\end{equation}
They compared their results against the ITDs reconstructed from global fits of the polarized gluon PDF from NNPDF~\cite{Nocera:2014gqa} and JAM~\cite{Zhou:2022wzm}, shown in Figure~\ref{fig:hadstruc_pol}.
They did not explicitly match the reconstructed ITDs to the RpITDs, arguing that the correction would be small compared with the current data precision.
The agreement between the lattice data and the NNPDF result is within one or two standard deviations for most of the data, as~can be seen in the left plot in Figure~\ref{fig:hadstruc_pol}.
JAM does not give a positivity constraint on $\Delta G$, finding that both positive and negative solutions are supported by the experimental data.
The ITD reproduced from the whole solution set is shown in the cyan band on the left, while the solutions that have $\Delta G>0$ are shown in green in both plots.
The $\Delta G <0$ solution set is plotted in maroon on the right.
This exercise shows that the RpITDs are in disagreement by several standard deviations with the $\Delta G <0$ solutions, while many of the data fall within one or two standard deviations of the $\Delta G >0$ solution.
This initially provided a strong suggestion that the true sign of the total gluon helicity should be positive.
However, this assertion was tested further in Ref.~\cite{Karpie:2023nyg}, and~those results are presented in the next section.
{It should be appreciated that all work in the gluon helicity case is still fairly exploratory and that systematics have not been explored in detail due to the large statistical noise and contamination term.
As available %EE: Please check meaning retained
statistics increase and methods to handle the contamination term evolve, there will be much left to explore here.}

\unskip

\section{Impacts on~Phenomenology}\label{sec:Impacts}
Though many systematics still need to be pinned down, as described in the previous section, there has already been substantial interest in adding lattice information to global fits, demonstrating that the combination of lattice and experimental data can constrain the PDFs more strongly than either one on their own~\cite{Lin:2017snn, Lin:2017stx, Bringewatt:2020ixn, JeffersonLabAngularMomentumJAM:2022aix, Gamberg:2022kdb, Hou:2022onq,Karpie:2023nyg, Hunt-Smith:2024khs,Ablat:2025xzm,Good:2025nny}.
These studies fall under a few different categories:
using the direct lattice $x$-space PDFs to constrain global fits, using the pseudo-PDF matching relations from Equation~(\ref{eq:pPDF-matching}) to treat the lattice data on equal footing with the experimental data, and~including information from lattice momentum fractions or charges.
As this paper focuses on pseudo-PDF and LaMET studies, we will restrict discussion to results for the first two methods with gluon PDF information.
The gluon PDF presents an excellent opportunity for lattice to contribute to global fits because the large-$x$ gluon PDF is quite poorly constrained {due to limited experimental constraints in this region.
This is in comparison to the isovector quark PDFs, where lattice studies of these PDFs have much better precision than those of the gluon PDF, but~experimental data constrains the quark PDFs to a level of precision with which lattice simply cannot compete.}
That is to say, the~potential impact is correspondingly greater in the gluon~case.

\begin{figure}[H]
    
    \includegraphics[width=0.5\linewidth]{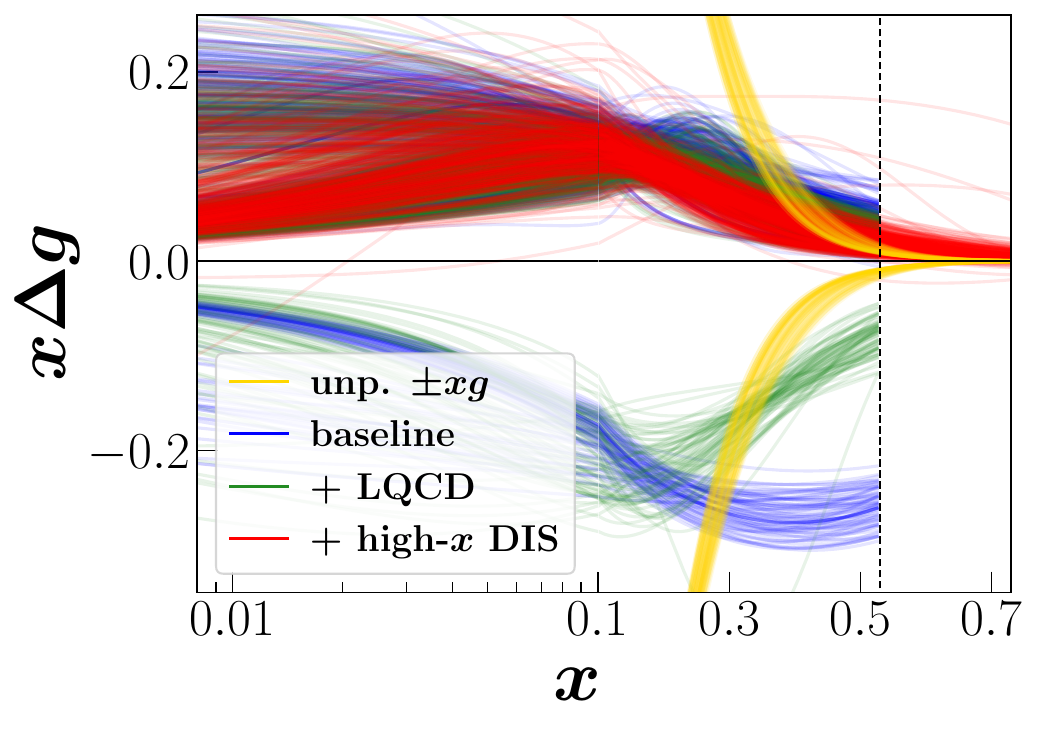}
    \caption{ {The} %MDPI: Please confirm whether an explanation of the dashed line needs to be added to the figure caption.%Author: Added explanation below
 results of the global fits of the unpolarized (yellow) and helicity (blue) gluon PDFs from JAM’s baseline fits, the~corresponding fits including lattice data (green), and~fits including both lattice and high-$x$ DIS data (red).
 {The dashed line represents a cutoff above $x\approx0.53$, where the baseline experimental data does not give constraints.}
 Figure taken from Ref.~\cite{Hunt-Smith:2024khs}.
    }
    \label{fig:delta_g_comp}
\end{figure}

In particular, the~results for the gluon helicity PDF from Ref.~\cite{Karpie:2023nyg} sparked immediate interest in adding this data into global fits to resolve the sign of $\Delta G$.
In Ref.~\cite{Karpie:2023nyg}, HadStruc %EE: Please check meaning retained
and JAM tackled this problem, following the methodology from Ref.~\cite{JeffersonLabAngularMomentumJAM:2022aix}, where RpITD data are included directly in the global fit and treated similarly to experimental data. 
This study found that, although~the initial experimental-data-only $\Delta G <0$ samples had poor agreement with the lattice RpITD, both $\Delta G <0$ and $\Delta G >0$ solutions could still provide acceptable fits, with~some modification to the shape of the $\Delta G <0$ solutions.
This study laid the groundwork for Ref.~\cite{Hunt-Smith:2024khs}, which incorporated additional high-$x$ deep inelastic scattering (DIS) data, finding that only combined lattice and experimental constraints disfavor $\Delta G <0$. 
The results of the fitted PDFs are shown in Figure~\ref{fig:delta_g_comp}.
This study should nevertheless be interpreted with caution, since the gluon RpITDs are obtained at finite lattice spacing and at a pion mass heavier than the physical value.
This provides strong motivation to obtain a physical-continuum gluon helicity PDF with high precision to test the robustness of these~conclusions. 

\begin{figure}[H]
    
    \includegraphics[width=0.45\linewidth]{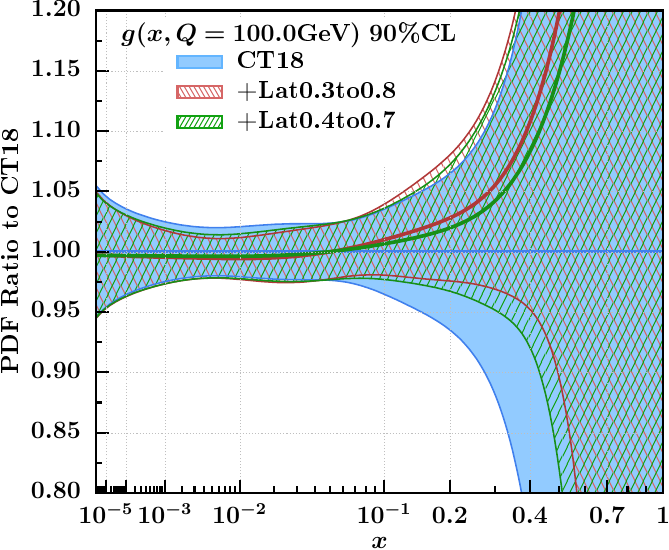}
    \includegraphics[width=0.45\linewidth]{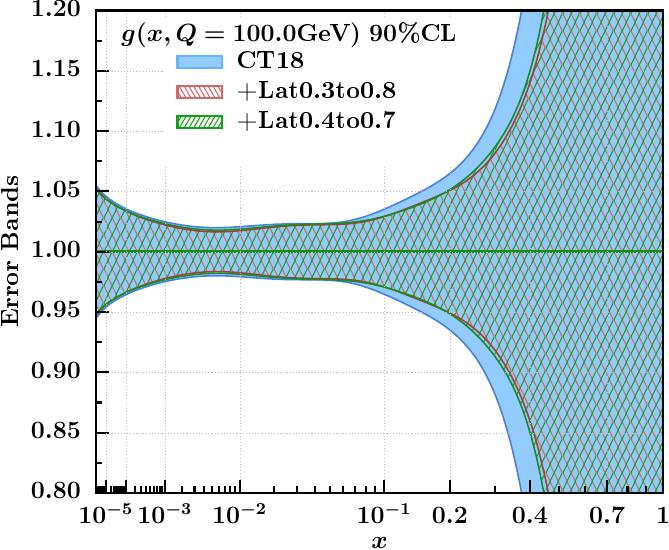}
    \caption{The central values (\textbf{left}) and uncertainty bands (\textbf{right}) of the gluon PDFs at $Q=100$~GeV from the CT18 fit alone (blue), as~well as from fits including additional lattice data in the ranges $x\in[0.3,0.8]$ (red) and $x\in[0.4,0.7]$ (green). The~uncertainty bands correspond to the 90\% confidence level. Figure taken from Ref.~\cite{Ablat:2025xzm}.}
    \label{fig:CT18_comp}
\end{figure}

In Ref.~\cite{Ablat:2025xzm}, MSULat and CTEQ followed Ref.~\cite{Hou:2022onq} to include unpolarized gluon PDF $x$-space information directly into the CT18 global fit. 
They used 100 points from the final PDFs in Ref.~\cite{Fan:2022kcb} in the ranges $x\in[0.3, 0.8]$ and $[0.4,0.7]$.
The first finding was that the physical-continuum-extrapolated gluon PDF had too much statistical noise to provide much constraint to the large-$x$ gluon PDF, so the $a\approx 0.09$ fm, $M_\pi \approx 310$ MeV results were used to provide an indication of how lattice could impact the unpolarized gluon PDFs.
The results are shown in Figure~\ref{fig:CT18_comp}.
The plot on the right shows the ratio of the new PDFs to the central value of the original CT18 gluon PDF. 
The lattice data seem to drag the large-$x$ PDF up slightly, but~these changes are statistically consistent with the original fit.
On the left plot, the~errors relative to each fit's central value are shown.
Both choices of $x$-range consistently give a small reduction in the large-$x$ relative error.
These results are promising but suggest that the lattice data still needs higher statistical precision to have a strong impact on global fits.
Additionally, the~methodology used in this study was originally proposed for LaMET PDFs, which give direct $x$-space information.
It would be interesting to see if using the RpITD data directly gives similar~results.

\begin{figure}[H]
    
    \includegraphics[width=0.45\linewidth]{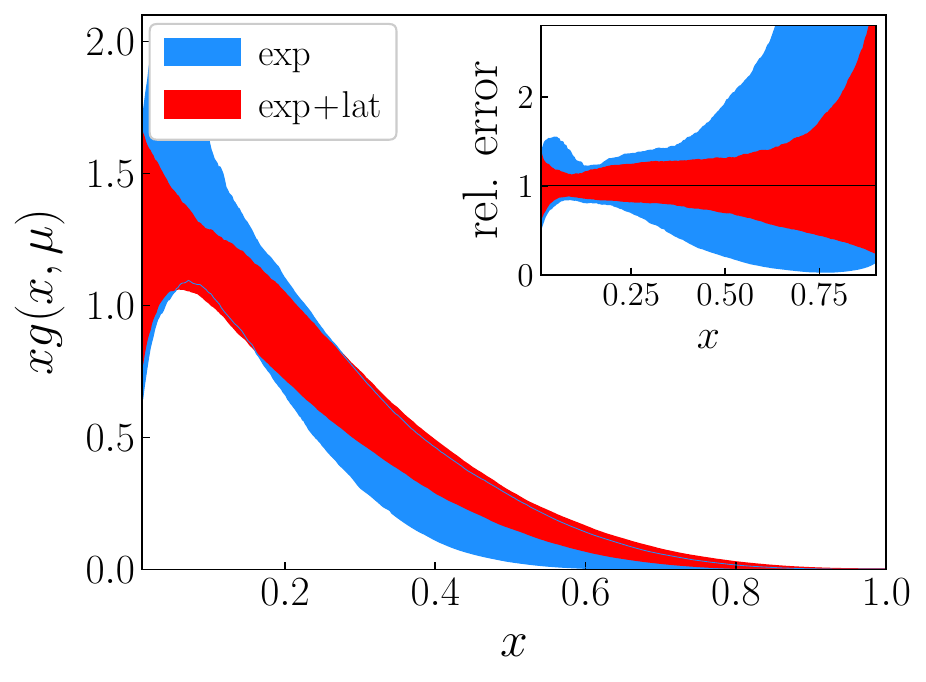}
    \includegraphics[width=0.45\linewidth]{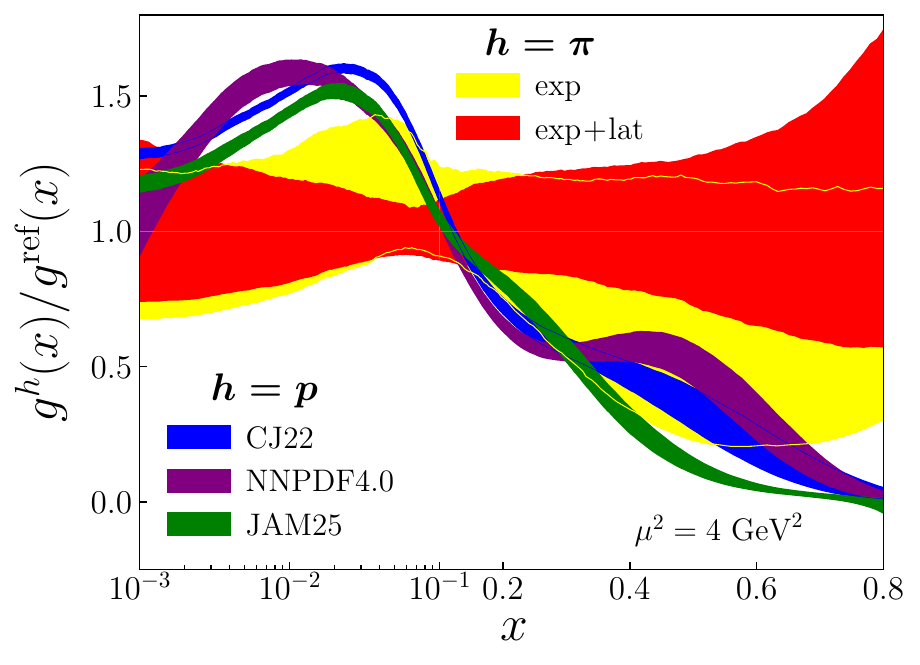}
    \caption{ (\textbf{Left}) Pion gluon PDFs at $\mu=m_c$ before (blue) and after (red) the inclusion of lattice data. The~inset shows the uncertainty bands relative to the median PDFs.
(\textbf{Right}) Gluon PDFs of the pion and proton, shown relative to the central value of the pion PDF fit. The~proton gluon PDFs are taken from CJ22~\cite{Cerutti:2025yji}, NNPDF4.0~\cite{NNPDF:2021njg}, and~JAM25~\cite{Cocuzza:2025qvf}. Figures are reproduced from Ref.~\cite{Good:2025nny}.}
    \label{fig:JAM_comp}
\end{figure}

Finally, the~pion gluon PDF is poorly constrained by experimental data, especially at large-$x$.
MSULat and JAM collaborated in Ref.~\cite{Good:2025nny} to include high-precision RpITD data from the $a\approx0.12 $ fm, $M_\pi \approx 310$ MeV ensemble directly into JAM's previous global fit for the pion PDFs~\cite{Barry:2021osv}, also following Ref.~\cite{JeffersonLabAngularMomentumJAM:2022aix}.
In this study, it was found that the inclusion of parameterizations for the higher-twist and discretization effects is important to fitting the RpITD data, improving the $Z$-score from 3 down to 1.
In either case, it could be seen that the lattice data have sensitivity to regions of $x$ that are complementary to those constrained by the experimental data, as~the $\chi^2_\text{exp.}$ was unaffected by fits with the inclusion of lattice data.
Two main results of this study are shown in Figure~\ref{fig:JAM_comp}.
On the right, it can be seen that the gluon PDF was strongly constrained at large-$x$ by the lattice data, but~is still consistent within one standard deviation of the experiment-only fits.
On the left, the~gluon PDF of the pion was compared with that of the proton.
An exciting result of this study, which warrants better understanding, is that the lattice data suggest that pion contains a significantly higher density of high-$x$ gluons than the proton, while both hadrons have a gluon momentum fraction of around $\langle x_g\rangle(\mu =2\text{ GeV}) \approx 0.4 $.
{This coincidence of momentum fractions could teach us something new about QCD, but~again,} this study should be extended using additional lattice ensembles to determine the continuum-extrapolated pion gluon PDF at the physical pion~mass.

\unskip

\section{Ongoing and New~Directions}\label{sec:new-direction}
Even with the early success in combining lattice gluon data with experimental measurements in phenomenological analyses, it must be emphasized that there is still substantial room for improvement in both statistical precision and control of systematic uncertainties.
On the pseudo-PDF side, a~major concern is the limited kinematic range of the lattice data.
This creates two issues.
First, the~inverse problem in the pseudo-PDF method is poorly constrained when the large-momentum data are too noisy.
The second is that the aforementioned $P_z$-dependent contamination in the helicity case remains significant in current kinematic ranges.
The direct way of handling this is to obtain higher-precision data at larger momentum; however, this is especially challenging in the gluon case.
Some efforts have been made using machine learning in Refs.~\cite{Khan:2022vot} and~\cite{Chowdhury:2024ymm}, focusing mainly on the helicity case.
These studies implement neural networks at several levels.
In the first study, a~theoretically motivated relation between RpITDs at different values of $P_z$ is combined with a neural network, used as a flexible parameterization to remove the $\mathcal{O}(M_H^2/P_z^2)$ contamination from the helicity RpITDs.
In the latter study, several random-forest-type machine learning algorithms are used to extrapolate the matrix elements to momenta beyond the current reach of lattice simulations, followed by a neural network parametrization to reconstruct the helicity and unpolarized PDFs.
It should be emphasized that this extrapolation method is complementary to direct calculations at higher momentum and, while conceptually similar to the long-distance extrapolation used in LaMET, is less directly motivated by the expected asymptotic behavior of the matrix elements.
To allow more flexible PDF parametrizations and better characterize the inverse problem in the pseudo-PDF method, Ref.~\cite{Kriesten:2025gti} explores using a variational autoencoder inverse mapper (VAIM) to learn multiple solutions to the inverse problem. 
A key result of this study is that the PDF extracted using this method from the same lattice data as in Ref.~\cite{Fan:2022kcb} has larger uncertainty than previously estimated using the rigid $x^\alpha (1-x)^\beta$ parameterization form, but~the results are still consistent with phenomenology.
The different methods for solving the inverse problem~\cite{Dutrieux:2024rem} should therefore be compared more systematically for noisy gluon~data.

Aside from these methods that are used later in the analysis pipeline, several algorithmic and theoretical methods have been proposed to improve the statistical precision of the lattice matrix elements themselves, all of which should be extensible to gluon PDFs.
First, as~discussed in Section~\ref{sec:theory}, new kinematically enhanced interpolators~\cite{Zhang:2025hyo,Reitinger:2026hta} have been proposed, which improve the statistical precision of the correlators at high momentum by factors of 10--50 compared to traditional interpolators, which would normally correspond to an increase in the number of measurements by a factor of $\mathcal{O}(100\text{--}1000)$.
{One of the new interpolators was used in Ref.~\cite{ChenChen:2025amm}, but~not all of the new interpolators have been fully explored in the gluon case.}
It has been recently shown for quark PDFs and transverse momentum distributions (TMDs) that fixing to the Coulomb gauge and removing the Wilson line from the operator definitions reduces noise, simplifies renormalization, and~results in consistent light-cone PDFs through the LaMET method~\cite{Gao:2023lny,Zhao:2023ptv,Gao:2024fbh,Bollweg:2024zet,Bollweg:2025iol,Bollweg:2025ecn, Grebe:2026qmt}.
During the drafting of this manuscript, there has been active work towards deriving the matching and collecting data to apply this Coulomb gauge method to the unpolarized gluon PDF.
Finally, exciting results have come recently showing that a machine learning technique called normalizing flows can be used to significantly improve the statistical precision of local gluonic observables~\cite{Abbott:2026ylv,Abbott:2026vui} by factors of order $\mathcal{O}(10\text{--}100)$ times.
{The authors of these studies describe normalizing flows and their applications to lattice QCD in much more detail in Refs.~\cite{Abbott:2023thq,Abbott:2024kfc,Albergo:2021vyo}. 
This is not a comprehensive list of their publications, but~the cited papers represent a few important summarizing works.
The last reference is especially pedagogical and provides tutorial code.}

\section{Conclusions and~Outlook}\label{sec:conclusion}
{In this review, we have highlighted the extensive effort to obtain the first results for gluon PDFs from the pseudo-PDF and LaMET methods and the attempts to address a range of systematic effects. 
The unpolarized nucleon gluon PDF has seen much of this progress, while the pion gluon PDF has trailed behind, and~the helicity PDF still has major signal-to-noise issues to tackle.}
The two theoretical methods each still face distinct challenges: the pseudo-PDF method is limited by an ill-conditioned inverse problem, while the LaMET expansion converges slowly for gluon PDFs. 
These challenges would benefit from high-precision data at larger hadron momentum, which will require finer lattices, increased statistics, and~algorithmic improvements.
It would be highly desirable to complete a systematic comparison of methods for solving the inverse problem, which include the basic PDF parameterization, machine learning~\cite{Chowdhury:2024ymm, Kriesten:2025gti}, and~Bayesian methods~\cite{Dutrieux:2024rem,Dutrieux:2025jed}.
Additionally, as~has been applied in the quark case, renormalization group resummation~\cite{Su:2022fiu}, leading renormalon resummation~\cite{Zhang:2023bxs}, and~threshold resummation~\cite{Ji:2023pba,Liu:2023onm,Ji:2024hit} should be explored to improve the convergence in the LaMET method, possibly alongside an NNLO matching calculation.
Despite these unresolved systematic uncertainties, several groups have already begun incorporating lattice QCD information into phenomenological analyses that need high-$x$ gluonic data.
These studies have mainly focused on the inclusion of pseudo-PDF data, which can be placed on an equal footing with experimental data.
This seems like a reasonable path forward, as~systematic effects in the lattice data can be parameterized and subsequently removed.
{No study has yet been completed that incorporates LaMET gluon-PDF results into global fits.}
One recently proposed approach for including lattice data into phenomenological extractions is the use of window observables~\cite{Karpie:2025ikc}, which is worth exploring in the LaMET case, as~long as systematics are carefully parameterized.
Finally, there is promising ongoing and future work that will improve the statistical precision of large-separation and high-momentum matrix elements by factors of $\mathcal{O}(10-100)$, as discussed in Section~\ref{sec:new-direction}.
{Overall, the~limited experimental constraints on the large-$x$ gluon presents a great opportunity for lattice to contribute. 
The goals of the next generation of unpolarized gluon PDF studies should be to develop a careful account of the error budget in these calculations and to approach a level of systematic and statistical precision sufficient to enable meaningful comparisons with, and~provide useful constraints on, phenomenological PDF determinations.}

\vspace{+6pt}

\authorcontributions{This review article was written, reviewed, and approved by both~authors. All authors have read and agreed to the published version.}

\funding{The work of WG and HL is supported by the US National Science Foundation under grant PHY~2514533.}

\dataavailability{No new data were created or analyzed for this review. Data sharing is not applicable to this~article.}

\acknowledgments{The authors thank Joseph Karpie, Joseph Delmar, Martha Constantinou, and~Chen Chen for sharing their final numerical results for compiling the comparison~plots.}

\conflictsofinterest{The authors declare no conflicts of~interest.}

\abbreviations{Abbreviations}{%
The following abbreviations are used in this manuscript:\\
\vspace{-11pt}

\noindent 
\begin{longtable}[l]{@{}ll}

{AMBER} %MDPI: Please confirm if the bold formatting is necessary; if not, please remove it. The following highlights are the same. %Author: Removed all bold
 & Apparatus for Meson and Baryon Experimental~Research\\

{DIS} & Deep inelastic~scattering\\

{EIC} & Electron-Ion~Collider\\

{ETMC} & Extended Twisted Mass~Collaboration\\

{GEVP} & Generalized eigenvalue~problem\\

{HISQ} & Highly improved staggered~quark\\

{HYP} & Hypercubic\\

{ITD} & Ioffe-time~distribution\\

{LaMET} & Large-momentum effective~theory\\

{LPC} & Lattice Parton~Collaboration\\

{LQCD} & Lattice quantum~chromodynamics\\

{NLO} & Next-to-leading~order\\

{NNLO} & Next-to-next-to-leading~order\\

{PDF} & Parton distribution~function\\

{QCD} & Quantum~chromodynamics\\

{RpITD} & Reduced pseudo-Ioffe-time~distribution\\

{TMD} &Transverse-momentum-dependent\\

{UV} &Ultraviolet
\end{longtable}
}

\begin{adjustwidth}{-\extralength}{0cm}
\reftitle{References}

\PublishersNote{}

\end{adjustwidth}

\end{document}